\documentclass{tlp}
\pdfoutput=1
\usepackage{aopmath}

\usepackage[latin1]{inputenc}
\usepackage[english]{babel}
\usepackage{latexsym}
\usepackage{graphicx}
\usepackage{subfigure}
\usepackage{algorithm}
\usepackage[noend]{algorithmic}
\usepackage{url}

\begin{document}
\bibliographystyle{acmtrans}

\hyphenation{know-ledge Santos making efficiently Kimmig using}

\long\def\comment#1{}

\graphicspath{{./figures/}}

\title{On the Implementation of the Probabilistic Logic Programming Language ProbLog}
\shorttitle{On the Implementation of ProbLog}

\author[A. KIMMIG et al]
{Angelika Kimmig, Bart Demoen and Luc De
Raedt \\ Departement Computerwetenschappen,
K.U. Leuven\\ Celestijnenlaan 200A - bus 2402, B-3001 Heverlee,
Belgium\\
\email{\{Angelika.Kimmig,Bart.Demoen,Luc.DeRaedt\}@cs.kuleuven.be}\\
\and V\'{\i}tor Santos Costa and Ricardo
  Rocha  \\ 
CRACS $\&$ INESC-Porto LA, Faculty of Sciences, University of Porto\\
R. do Campo Alegre 1021/1055, 4169-007 Porto, Portugal\\
\email{\{vsc,ricroc\}@dcc.fc.up.pt}\\ }

\setcounter{page}{1}

\maketitle

\setcounter{footnote}{0}
\begin{abstract}
  The past few years have seen a surge of interest in the field of
  probabilistic logic learning and statistical relational learning.  In
  this endeavor, many probabilistic logics have been
  developed. ProbLog is a recent probabilistic extension of Prolog
  motivated by the mining of large biological networks. In ProbLog,
facts can be labeled with probabilities. These facts are treated as mutually independent random variables that indicate whether these facts belong to a randomly sampled program.
 Different kinds of queries can be posed to ProbLog programs. We
  introduce algorithms that allow the efficient execution
  of these queries, discuss their implementation on top of the
  YAP-Prolog system,  and evaluate their performance in the context of
  large networks of biological entities.\\
\emph{To appear in Theory and Practice of Logic Programming (TPLP)}
\end{abstract}

\section{Introduction}\label{sec:intro}
In the past few years, a multitude of different 
formalisms combining probabilistic reasoning with logics, databases, or logic programming  
has been developed. 
Prominent examples include PHA and ICL~\cite{Poole:93,Poole00},
PRISM~\cite{SatoKameya:01}, SLPs~\cite{Muggleton96},
ProbView~\cite{Lakshmanan}, CLP($\cal BN$)~\cite{Costa03:uai},
CP-logic~\cite{Vennekens}, Trio~\cite{Trio}, probabilistic
Datalog~(pD)~\cite{Fuhr00}, and probabilistic databases~\cite{DalviS04}.  Although these logics have been
traditionally studied in the knowledge representation and database
communities, the focus is now often on a machine learning
perspective, which imposes new requirements. 
First, these logics must be
simple enough to be learnable and at the same time sufficiently expressive to
support interesting probabilistic inferences.  Second, because
learning is computationally expensive and requires answering long
sequences of possibly complex queries, inference in such logics must
be fast, although inference in even the simplest probabilistic logics is computationally hard.

In this paper, we study these problems in the context of a simple
probabilistic logic,  ProbLog~\cite{DeRaedt07}, which has been used for
learning in the context of large
biological networks where edges are labeled with probabilities.  Large
and complex networks of biological concepts (genes, proteins,
phenotypes, etc.)  can be extracted from public databases, and
probabilistic links between concepts can be obtained by various
techniques~\cite{Sevon06}. 
ProbLog is essentially an extension of Prolog where a program defines a distribution over all  its possible  non-probabilistic subprograms. Facts are labeled with probabilities and treated as mutually independent random variables indicating whether or not the corresponding fact belongs to a randomly sampled program.
The success
probability of a query is defined as the probability that it succeeds
in such a random subprogram. The semantics of ProbLog is not new: it is an instance of the distribution semantics~\cite{Sato:95}. This is a
well-known semantics for probabilistic logics that has been
(re)defined multiple times in the literature, often in a more limited database setting; cf.~\cite{Dantsin,Poole:93,Fuhr00,Poole00,DalviS04}. Sato has, however, shown that the semantics is also well-defined in the case of a countably infinite set of random variables and formalized it in his well-known distribution semantics~\cite{Sato:95}.
However, even though relying on the same semantics, in order to allow efficient inference, systems such as PRISM~\cite{SatoKameya:01} and PHA~\cite{Poole:93} additionally require all proofs of a query to be mutually exclusive. Thus, they cannot easily represent the type of network analysis tasks that motivated ProbLog. ICL~\cite{Poole00} extends PHA to the case where proofs need not be mutually exclusive. In contrast to the ProbLog implementation presented here, Poole's AILog2, an implementation of ICL, uses a meta-interpreter and is not tightly integrated with Prolog. 

We contribute exact and approximate inference algorithms for
ProbLog.  We present algorithms for computing the success and
explanation probabilities of a query, and show how they can be 
efficiently implemented combining Prolog
inference with Binary Decision Diagrams (BDDs)~\cite{Bryant86}. In addition to an iterative deepening algorithm that computes an approximation along the lines of~\cite{Poole93:jrnl}, we further
adapt the Monte Carlo approach 
used by~\cite{Sevon06} in the context of biological network inference. These two approximation algorithms compute an upper and a lower bound on the success probability. We
also contribute an additional approximation algorithm that computes a lower bound using
only the $k$ most likely proofs.

The key contribution of this paper is the tight integration of these
algorithms in the state-of-the-art 
YAP-Prolog system. This integration includes several improvements over the initial implementation used in~\cite{DeRaedt07}, which are needed to use ProbLog to effectively query Sevon's Biomine
network~\cite{Sevon06} containing about 1,000,000 nodes and
6,000,000 edges, as will be shown in the experiments.

This paper is organised as follows. After introducing ProbLog and its semantics in Section 2,
we present several algorithms for exact and approximate inference in Section 3. Section 4 then
discusses how these algorithms are implemented in YAP-Prolog, and Section 5 reports on experiments
that validate the approach. Finally, Section 6 concludes and touches upon related work.

\section{ProbLog}\label{sec:problog}
A ProbLog program consists of a set of labeled facts $p_i::c_i$ together with a set of definite clauses. Each ground instance (that is, each instance not containing variables) of such a fact $c_i$ is true with probability $p_i$, that is, these facts correspond to random variables. We assume that these variables are mutually independent.\footnote{If the program contains multiple instances of the same fact, they correspond to different random variables, i.e.~$\{p::c\}$ and $\{p::c, p::c\}$ are different ProbLog programs.}
The definite clauses allow one to add arbitrary \emph{background knowledge} (BK).

\begin{figure}[t]
\centering
\includegraphics[]{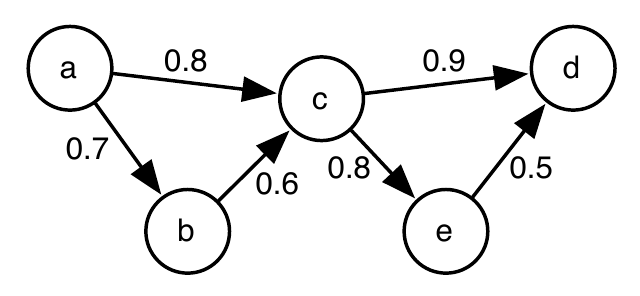}
\caption{ Example of a probabilistic graph: edge labels
 indicate the probability that the edge is part of the graph.}\label{fig:Ex}
\end{figure}

Figure~\ref{fig:Ex} shows
a small probabilistic graph that we shall use as running example in the text.
It can be encoded in ProbLog as follows:
\begin{equation}
\begin{array}{lllll}
0\ldotp8 :: \mathtt{edge(a,c)\ldotp} & ~~~~ &  0\ldotp7 :: \mathtt{edge(a,b)\ldotp} & ~~~~ &  0\ldotp8 :: \mathtt{edge(c,e)\ldotp} \\
0\ldotp6 :: \mathtt{edge(b,c)\ldotp} & ~~~~ &  0\ldotp9 :: \mathtt{edge(c,d)\ldotp} & ~~~~  &  0\ldotp5 :: \mathtt{edge(e,d)\ldotp} 
\end{array}
\end{equation}
Such a probabilistic graph can be used to sample subgraphs by tossing
a coin for each edge.  
Given a ProbLog program $T=\{p_1::c_1,\cdots,p_n::c_n\} \cup BK$ and a finite set of possible substitutions $\{\theta_{j1}, \ldots \theta_{ji_j}\}$ for each probabilistic fact $p_j::c_j$, let $L_T$ denote the maximal set of \emph{logical} facts that can be added to $BK$, that is,  $L_T=\{c_1\theta_{11}, \ldots , c_1\theta_{1i_1}, \cdots, c_n\theta_{n1}, \ldots , c_n\theta_{ni_n}\}$. As the random variables corresponding to facts in $L_T$ are mutually independent, the ProbLog program defines a probability distribution over ground logic programs $L \subseteq L_T$:
\begin{equation}
    P(L|T)=\prod\nolimits_{c_i\theta_j\in L}p_i\prod\nolimits_{c_i\theta_j\in L_T\backslash L}(1-p_i)\ldotp
\end{equation}
Since the background knowledge $BK$ is fixed and there is a one-to-one mapping between ground definite clause programs and Herbrand interpretations, a ProbLog program thus also defines a distribution over its Herbrand interpretations. 
Sato has shown how this semantics can be generalized to the countably infinite case; we refer to~\cite{Sato:95} for details. For ease of readability, in the remainder of this paper we will restrict ourselves to the finite case and assume all probabilistic facts in a ProbLog program to be ground.
We extend our example with the following background knowledge:
\begin{equation}
\begin{array}{lll}
\mathtt{path(X,Y)} &  \mathtt{:-} &  \mathtt{edge(X,Y)\ldotp} \\
\mathtt{path(X,Y)} &  \mathtt{:-} &  \mathtt{edge(X,Z), path(Z,Y)\ldotp} 
\end{array}
\end{equation}
We can then ask for the probability that there exists a path
between two nodes, say \emph{c} and \emph{d}, in our probabilistic graph, that is, we query for the probability that a randomly sampled subgraph contains the
edge from \emph{c} to \emph{d}, or the path from \emph{c} to \emph{d}
via \emph{e} (or both of these).
Formally, the \emph{success probability} $P_s(q|T)$ of a query $q$ in
a ProbLog~program~$T$ is the marginal of $P(L|T)$ with respect to $q$, i.e.
\begin{equation}
P_s(q|T) = \sum\nolimits_{L\subseteq L_T}P(q|L)\cdot P(L|T)\;, \label{eq:p_suc}
\end{equation}
where $P(q|L) = 1$ if there exists a $\theta$ such that $L\cup BK\models
q\theta$, and $P(q|L)=0$ otherwise.  In other words, the success
probability of query $q$ is the probability that the query $q$ is
\emph{provable} in a randomly sampled logic program.

In our example, $40$ of the $64$ possible subprograms allow one to prove \emph{path$(c,d)$}, namely all those that contain at least the edge from \emph{c} to \emph{d} or both the edge from
\emph{c} to \emph{e} and from \emph{e} to \emph{d}, so the success probability of that query is the sum of the probabilities of these programs: 
$P_s(path(c,d)|T)=P(\{ab,ac,bc,cd,ce,ed\}|T)+\ldots +P(\{cd\}|T)=0\ldotp94$, where $xy$ is used 
as a shortcut for \emph{edge$(x,y)$} when listing elements of a subprogram. We will use this convention throughout the paper. Clearly, listing all subprograms is infeasible in practice; an alternative approach will be discussed in Section~\ref{sec:exact}. 

A ProbLog program also defines the probability of a \emph{specific} proof $E$, also called \emph{explanation}, of some query $q$, which is again a marginal of $P(L|T)$. Here, an \emph{explanation} is a minimal subset of the probabilistic facts that together with the background knowledge entails $q\theta$ for some substitution $\theta$. Thus, the probability of such an explanation $E$ is that of sampling a logic program $L\cup E$ that contains at least all the probabilistic facts in $E$, that is, the marginal with respect to these facts:
\begin{equation}
P(E|T)  =   \sum\nolimits_{ L\subseteq (L_T\backslash E)} P(L\cup E |T) = \prod\nolimits_{c_i \in E}p_i\label{eq:deriv_px}
\end{equation}
The \emph{explanation probability} $P_x(q|T)$ is then defined
as the probability of the most likely explanation or proof of the
query~$q$
\begin{equation}
P_x(q|T)  =  \max\nolimits_{E\in E(q)}P(E|T)
=   \max\nolimits_{E\in E(q)} \prod_{c_i \in E}p_i,\label{eq:p_exp}
\end{equation}
where $E(q)$ is the set of all explanations for query
$q$, i.e., all minimal sets $E\subseteq L_T$ of probabilistic facts such that $E \cup BK \models q$~\cite{Kimmig07}.

In our example, the set of all explanations for \emph{path$(c,d)$}
contains the edge from \emph{c} to \emph{d} (with
probability 0.9) as well as the path consisting of the edges from
\emph{c} to \emph{e} and from \emph{e} to \emph{d} (with probability
$0\ldotp8\cdot 0\ldotp5=0\ldotp4$). Thus, $P_x(path(c,d)|T)=0\ldotp9$.

The ProbLog semantics is essentially a distribution
semantics~\cite{Sato:95}.  Sato has rigorously shown that this class
of programs defines a joint probability distribution over the set of
possible least Herbrand models of the program (allowing functors), that is, of the
background knowledge $BK$ together with a
subprogram $L \subseteq L_T$; for further details we refer to~\cite{Sato:95}. The distribution semantics has been used widely in the
literature, though often under other names or in a more restricted setting; see e.g.~\cite{Dantsin,Poole:93,Fuhr00,Poole00,DalviS04}.

\section{Inference in ProbLog}\label{sec:inference}
This section discusses algorithms for computing exactly or
approximately the success and explanation probabilities of ProbLog
queries. It additionally contributes a new algorithm for Monte Carlo approximation of success probabilities.

\subsection{Exact Inference}\label{sec:exact}
Calculating the \emph{success probability} of a query using
Equation~(\ref{eq:p_suc}) directly is infeasible for all but the tiniest
programs, as the number of subprograms to be checked is exponential in the number of probabilistic facts. 
However, as we have seen in our example in Section~\ref{sec:problog}, we can describe all subprograms allowing for a specific proof by means of the facts that such a program has to contain, i.e., all the ground probabilistic facts used in that proof. As probabilistic facts correspond to random variables indicating the presence of facts in a sampled program, we alternatively  denote proofs by conjunctions of such random variables. 
In our example, query \emph{path(c,d)} has two proofs in the full program: \emph{\{edge(c,d)\}} and \emph{\{edge(c,e),edge(e,d)\}}, or, using logical notation, $cd$ and $ce \wedge ed$. The set of all subprograms containing \emph{some} proof thus can be described by a disjunction over all possible proofs, in our case, $cd \vee (ce \wedge ed)$. 
This idea forms the basis for the inference method presented in~\cite{DeRaedt07}, which uses two steps:
\begin{enumerate}
\item Compute the proofs of the query $q$ in the logical
part of the theory $T$, that is, in $BK \cup L_T$. The result will be a DNF
formula.
\item Compute the probability of this formula.
\end{enumerate}
Similar approaches are used for PRISM~\cite{SatoKameya:01}, ICL~\cite{Poole00} and pD~\cite{Fuhr00}.

The probability of a single given proof, cf.~Equation~(\ref{eq:deriv_px}), is the marginal over all programs allowing for that proof, and thus equals the product of the probabilities of the facts used by that proof.  However, we cannot directly sum the results for the different proofs to obtain the success probability, as a specific subprogram can allow several proofs and therefore contributes to the probability of each of these proofs. Indeed, in our example, all programs that are supersets of \emph{\{edge(c,e),edge(e,d),edge(c,d)\}} contribute to the marginals of both proofs and would therefore be counted twice if summing the probabilities of the proofs. However, for mutually exclusive conjunctions, that is, conjunctions describing disjoint sets of subprograms, the probability is the sum of the individual probabilities.
This situation can be achieved by adding \emph{negated} random variables to a conjunction, thereby explicitly excluding subprograms covered by another part of the formula  from the corresponding part of the sum. 
 In the example, extending $ce \wedge ed$ to $ce \wedge ed \wedge \neg cd$ reduces the second part of the sum to those programs not covered by the first: 
\[P_s(path(c,d)|T)=P(cd \vee (ce\wedge ed)|T)\]\[= P(cd|T)+P(ce\wedge ed\wedge\neg cd|T)  \]\[= 0\ldotp9 + 0\ldotp8\cdot0\ldotp5\cdot(1-0\ldotp9)=0\ldotp94\]

\noindent However, as the number of proofs grows, disjoining them gets more involved. Consider for example the query \emph{path(a,d)} which has four different but highly interconnected proofs. In general, this problem is known as the \emph{disjoint-sum-problem} or the two-terminal network reliability problem, which is \#P-complete~\cite{Valiant1979}.

Before returning to possible approaches to tackle the disjoint-sum-problem at the end of this section, we will now discuss the two steps of ProbLog's exact inference in more detail.

\begin{figure}[t]
\centering
\includegraphics[scale=0.7]{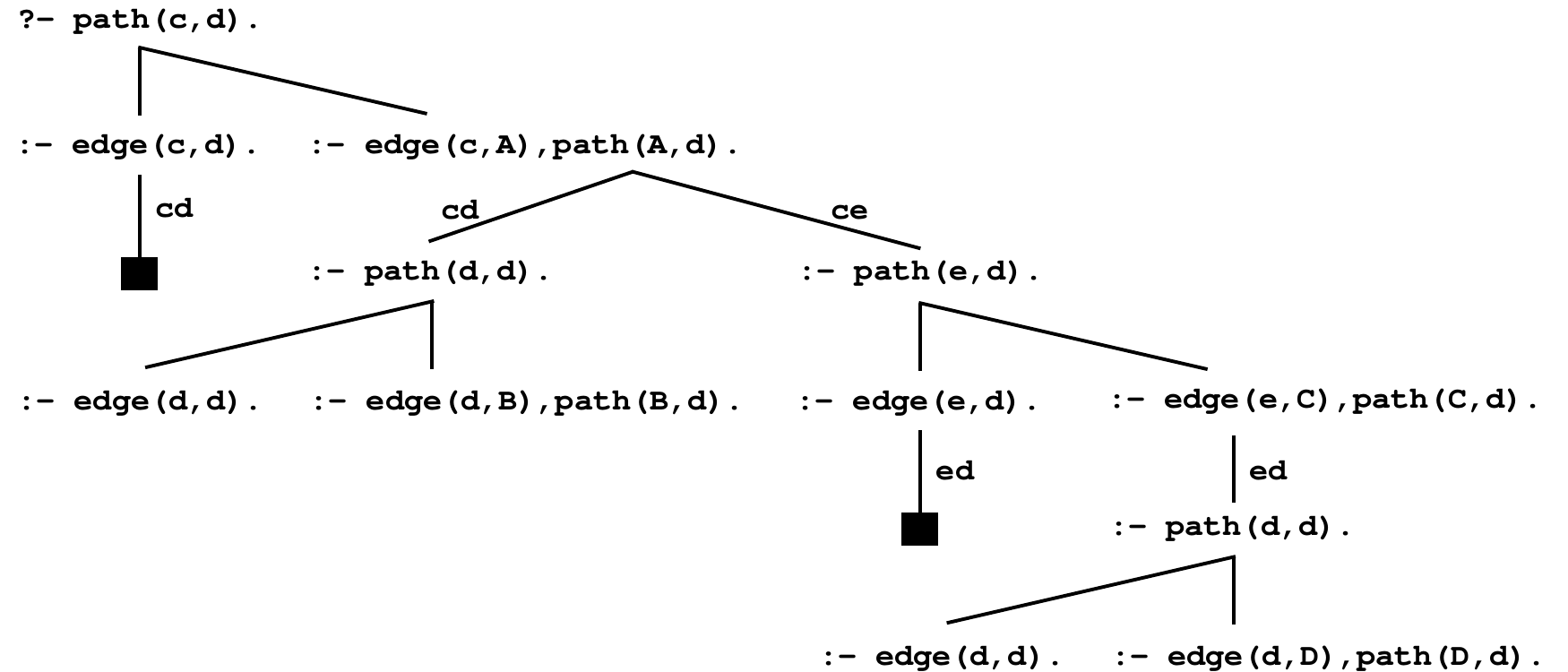}
\caption{SLD-tree for query \emph{path$(c,d)$.}}
\label{fig:SLD}
\end{figure}
Following Prolog, the first step employs SLD-resolution to obtain all different
proofs. As an example, the SLD-tree for the query \emph{?- path$(c,d)$.}
is depicted in Figure~\ref{fig:SLD}.
Each successful proof in the SLD-tree uses a set of ground probabilistic facts $\{p_1::c_1, \cdots, p_k::c_k\}
\subseteq T$.  These facts
are necessary for the proof, and the proof is \emph{independent} of other
probabilistic facts in~$T$.  

Let us now introduce a Boolean random variable $b_i$ for each ground probabilistic fact
$p_i::c_i \in T$, indicating whether $c_i$ is in a sampled logic program, that is,
$b_i$ has probability $p_i$ of being true.\footnote{For better readability, we do not write substitutions explicitly here.}  
A particular proof of query $q$ involving ground facts $\{p_1::c_1, \cdots, p_k::c_k\}
\subseteq T$ is thus represented by the conjunctive formula $b_1
\wedge \cdots \wedge b_k$, which at the same time represents the set of all subprograms containing these facts.  
Furthermore, using $E(q)$ to denote the set of proofs or explanations of the goal $q$, the set of all subprograms containing \emph{some} proof of $q$ can be denoted by $\bigvee_{e \in E(q) } \, \bigwedge_{c_i \in e} b_i $, as the following derivation shows:
\begin{eqnarray*}
\bigvee_{e \in E(q) } \, \bigwedge_{c_i \in e} b_i & = & \bigvee_{e \in E(q) } \left( \bigwedge_{c_i \in e} b_i \wedge \bigwedge_{c_i \in L_T \backslash e} (b_i \vee \neg b_i)\right)\\
& = & \bigvee_{e \in E(q) } \bigvee_{L \subseteq L_T\backslash e }  \left( \bigwedge_{c_i \in e} b_i \wedge \left(\bigwedge_{c_i \in L} b_i  \wedge \bigwedge_{c_i \in L_T \backslash (L\union e)} \neg b_i\right)\right)\\
& = & \bigvee_{e \in E(q) , L \subseteq L_T\backslash e }  \left( \bigwedge_{c_i \in  L \union e} b_i  \wedge \bigwedge_{c_i \in L_T \backslash (L\union e)} \neg b_i\right)\\
& = & \bigvee_{ L \subseteq L_T, \exists\theta L\union BK \models q\theta }  \left( \bigwedge_{c_i \in  L } b_i  \wedge \bigwedge_{c_i \in L_T \backslash L} \neg b_i\right)
\end{eqnarray*}
We first add all possible ways of extending a proof $e$ to a full sampled program by considering each fact not in $e$ in turn. We then note that the disjunction of these fact-wise extensions can be written on the basis of sets. Finally, we rewrite the condition of the disjunction in the terms of Equation~(\ref{eq:p_suc}). This is possible as each subprogram that is an extension of an explanation of $q$ entails some ground instance of $q$, and vice versa, each subprogram entailing $q$ is an extension of some explanation of $q$. 
As the DNF now contains conjunctions representing fully specified programs, its probability is a sum of products, which directly corresponds to Equation~(\ref{eq:p_suc}):
\begin{eqnarray*}
\lefteqn{ P(\bigvee_{ L \subseteq L_T, \exists\theta L\union BK \models q\theta }  \left( \bigwedge_{c_i \in  L } b_i  \wedge \bigwedge_{c_i \in L_T \backslash L} \neg b_i\right)) }\\
&=& \sum_{ L \subseteq L_T, \exists\theta L\union BK \models q\theta }  \left( \prod_{c_i \in  L } p_i  \cdot \prod_{c_i \in L_T \backslash L} (1- p_i)\right)\\
&=& \sum_{ L \subseteq L_T, \exists\theta L\union BK \models q\theta }  P(L|T)
\end{eqnarray*}
We thus obtain the following alternative characterisation of the success probability:
\begin{equation}
  P_s(q|T) = P\left( \bigvee_{e \in E(q) } \, \bigwedge_{c_i \in e} b_i \right)
  \label{eq:dnf}
\end{equation}
where $E(q)$ denotes the set of proofs or explanations of the goal $q$
and $b_i$ denotes the Boolean variable corresponding to ground probabilistic 
fact $p_i::c_i$.  Thus, the problem of computing the
success probability of a ProbLog query can be reduced to that of
computing the probability of a DNF formula.

However, as argued above, due to overlap between different conjunctions, the proof-based DNF of Equation~(\ref{eq:dnf}) cannot directly be transformed into a sum of products. 
Computing the probability of DNF formulae thus involves solving the disjoint-sum-problem, and therefore  is itself a \#P-hard problem. Various
algorithms have been developed to tackle this problem. The pD-engine HySpirit~\cite{Fuhr00} uses the inclusion-exclusion principle, which is reported to scale to about ten proofs. For ICL, which extends PHA by allowing non-disjoint proofs, \cite{Poole00} proposes a symbolic disjoining algorithm, but does not report scalability results. 
Our implementation of ProbLog employs Binary
Decision Diagrams (BDDs)~\cite{Bryant86}, an efficient graphical
representation of a Boolean function over a set of variables, which scales to tens of thousands of proofs; see
Section~\ref{sec:BDD} for more details. PRISM~\cite{SatoKameya:01} and PHA~\cite{Poole:93} differ from the systems mentioned above in that they avoid the disjoint-sum-problem by requiring the user to write programs such that proofs are guaranteed to be disjoint. 

On the other hand, as the \emph{explanation probability} $P_x$ exclusively depends on the probabilistic facts used in one proof, it can be calculated using a simple branch-and-bound approach based on the SLD-tree, where partial proofs are discarded if their probability drops below that of the best proof found so far.

\subsection{Approximative Inference}
As the size of the DNF formula grows with the number of proofs, its
evaluation can become quite expensive, and ultimately infeasible. For
instance, when searching for paths in graphs or networks, even in
small networks with a few dozen edges there are easily $O(10^6)$
possible paths between two nodes. ProbLog therefore includes several
approximation methods.

\subsubsection{Bounded Approximation}
The first approximation algorithm, a slight variant of the one proposed in~\cite{DeRaedt07}, uses DNF formulae to obtain both an upper and a lower bound on the probability of a query. It is closely related to work by~\cite{Poole93:jrnl} in the context of PHA, but adapted towards ProbLog.  The method relies on two observations.

First, we remark that the DNF formula describing sets of proofs is \emph{monotone}, meaning that adding more proofs will never decrease the probability of the formula being true. Thus, formulae describing subsets of the full set of proofs of a query will always give a lower bound on the query's success probability. In our example, the lower bound obtained from the shorter proof would be $P(cd|T) = 0\ldotp9$, while that from the longer one would be $P(ce\wedge ed|T) = 0\ldotp4$.

Our second observation is that the probability of a proof $b_1 \wedge \ldots\wedge b_n$ will always be at most the probability of an arbitrary prefix $b_1 \wedge \ldots\wedge b_i, i\leq n$. 
In our example, the probability of the second proof will be at most the probability of its first edge from $c$ to $e$, i.e., $P(ce|T) = 0\ldotp8 \geq 0\ldotp4$. As disjoining sets of proofs, i.e., including information on facts that are \emph{not} elements of the subprograms described by a certain proof, can only decrease the contribution of single proofs, this upper bound carries over to a set of proofs or partial proofs, as long as prefixes for all possible proofs are included. Such sets can be obtained from an incomplete SLD-tree, i.e., an SLD-tree where branches are only extended up to a certain point.

This motivates ProbLog's \emph{bounded approximation algorithm}. The algorithm relies on a probability threshold $\gamma$ to stop growing the SLD-tree and thus obtain DNF formulae for the two bounds\footnote{Using a probability threshold instead of the depth bound of~\cite{DeRaedt07} has been found to speed up convergence, as upper bounds have been found to be tighter on initial levels.}.  The lower bound formula $d_1$ represents all proofs with a probability above the current threshold. The upper bound formula $d_2$ additionally includes all derivations that have been stopped due to reaching the threshold, as these still \emph{may} succeed. Our goal is therefore to grow $d_1$ and $d_2$ in order to decrease $P(d_2|T)-P(d_1|T)$.

Given an acceptance threshold $\delta_p$, an initial probability threshold $\gamma$, and a shrinking factor $\beta\in(0,1)$, the algorithm proceeds in an iterative-deepening manner as outlined in Algorithm~\ref{alg:delta}. Initially, both $d_1$ and $d_2$ are set to \textsc{False}, the neutral element with respect to disjunction, and the probability bounds are $0$ and $1$, as we have no full proofs yet, and the empty partial proof holds in any model.
\begin{algorithm}[t]
  \caption{Bounded approximation using iterative deepening with probability thresholds.}
\label{alg:delta}
\begin{algorithmic}
\FUNCTION{\textsc{Bounds}(interval width $\delta_p$, initial threshold $\gamma$, constant $\beta\in(0,1)$)}
\STATE $d_1 = $ \textsc{False}; $d_2 = $ \textsc{False};  $P(d_1|T) =0$; $P(d_2|T) = 1$; 
\WHILE{$P(d_2|T) - P(d_1|T)>\delta_p$}
\STATE $p = $\textsc{True};
\REPEAT
\STATE Expand current proof $p$
\UNTIL {either $p$:
    \STATE $\quad \quad$ (a) Fails, in this case backtrack to the remaining choice points;
    \STATE $\quad \quad$ (b) Succeeds, in this case set $d_1 := d_1 \vee p$ and $d_2 := d_2 \vee p$;
    \STATE $\quad \quad$ (c) $P(p|T) < \gamma$, in this case set $d_2 := d_2 \vee p$}
\IF{$d_2 == $ \textsc{False}}  \STATE set $d_2 = $ \textsc{True} \ENDIF
\STATE Compute $P(d_1|T)$ and $P(d_2|T)$
\STATE $\gamma := \gamma\cdot\beta$  
\ENDWHILE
\STATE return $[P(d_1|T),P(d_2|T)]$
\ENDFUNCTION
\end{algorithmic}
\end{algorithm} 

It should be clear that $P(d_1|T)$ monotonically increases, as the number of proofs never decreases. On the other hand, as explained above, if $d_2$ changes from one iteration to the next, this is always because a partial proof $p$ is either removed from $d_2$  and therefore no longer contributes to the probability, or it is replaced by proofs $p_1,\ldots , p_n$, such that $p_i = p \land s_i$, hence $P(p_1 \lor \ldots \lor p_n|T) = P(p \land s_1\lor\ldots\lor p\land s_n|T) = P(p \land ( s_1\lor\ldots\lor s_n)|T)$. As proofs are subsets of the probabilistic facts in the ProbLog program, each literal's random variable appears at most once in the conjunction representing a proof, even if the corresponding subgoal is called multiple times when constructing the proof. 
We therefore know that the literals in the prefix $p$ cannot be in any suffix $s_i$, hence, given ProbLog's independence assumption, $P(p \land ( s_1\lor\ldots\lor s_n)|T) = P(p|T)P(s_1\lor\ldots\lor s_n|T) \leq P(p|T)$. Therefore, $P(d_2)$ monotonically decreases.

As an illustration, consider a probability threshold  $\gamma =0\ldotp9$ for the
SLD-tree in Figure~\ref{fig:SLD}. In this case, $d_1$ encodes the left
success path while $d_2$ additionally encodes the path up to
\emph{path$(e,d)$}, i.e., $d_1 = cd$ and $d_2 = cd \vee ce$, whereas the
formula for the full SLD-tree is $d = cd \vee (ce \wedge ed)$. The lower bound thus is $0\ldotp9$, the upper bound (obtained by disjoining $d_2$ to $cd \vee (ce\wedge\neg cd)$)  is $0\ldotp98$, whereas the true probability is $0\ldotp94$.

Notice that in order to implement this algorithm we need to compute the probability of a set of proofs. This task will be described in detail in Section~\ref{sec:implementation}.

\subsubsection{K-Best} Using a fixed number of proofs to approximate the
probability allows better control of the overall complexity, which is crucial if large numbers of queries have to be evaluated, e.g., in the context of parameter learning. 
\cite{Gutmann08}~therefore introduces the $k$-probability $P_k(q|T)$, which approximates
the success probability by using the $k$-best (that is, the $k$ most likely)
explanations instead of all proofs when building the DNF formula used
in Equation~(\ref{eq:dnf}):
\begin{equation}
P_k(q|T)  = P\left( \bigvee_{e \in E_k(q) } \, \bigwedge_{b_i \in var(e)} b_i \right)\label{eq:p_k}
\end{equation}
where $E_k(q)=\{e \in E(q)|P_x(e)\geq P_x(e_k)\}$ with $e_k$ the $k$th
element of $E(q)$ sorted by non-increasing probability.  Setting
$k=\infty$ leads to the success probability, whereas $k=1$ corresponds to the explanation probability provided that there is a single best proof.
The branch-and-bound approach used to calculate the explanation probability can directly be generalized to finding the $k$-best proofs; cf. also~\cite{Poole:93}.

To illustrate $k$-probability, we consider again our example graph,
but this time with query \emph{path$(a,d)$}. This query has four proofs,
represented by the conjunctions $ac\wedge cd$, $ab\wedge bc \wedge
cd$, $ac\wedge ce \wedge ed$ and $ab\wedge bc \wedge ce \wedge ed$,
with probabilities $0\ldotp72$, $0\ldotp378$, $0\ldotp32$ and $0\ldotp168$
respectively. As $P_1$ corresponds to the explanation probability
$P_x$, we obtain $P_1(path(a,d))=0\ldotp72$. For $k=2$, the overlap between the
best two proofs has to be taken into account: the second proof only
adds information if the first one is absent. As they share edge
$cd$, this means that edge $ac$ has to be missing, leading to
$P_2(path(a,d))=P((ac\wedge cd) \vee (\neg ac \wedge ab\wedge bc
\wedge cd))=0\ldotp72+(1-0\ldotp8)\cdot 0\ldotp378=0\ldotp7956$. Similarly, we obtain
$P_3(path(a,d))=0\ldotp8276$ and $P_k(path(a,d))=0\ldotp83096$ for $k\geq 4$.

\subsubsection{Monte Carlo}\label{sec:mc_method}
As an alternative approximation technique, we propose a Monte Carlo method, where we proceed as follows.

\noindent Execute until convergence:
  \begin{enumerate}
  \item Sample a logic program from the ProbLog program
  \item Check for the existence of some proof of the query of interest
  \item Estimate  the query probability $P$ as the fraction of samples where the query is provable
  \end{enumerate}
We estimate convergence by computing the 95\% confidence interval at each $m$ samples. Given a large number $N$ of samples, we  can use the standard normal approximation interval to the binomial distribution:

\[ \delta \approx 2\times\sqrt{\frac{P.(P-1)}{N}} \]

\noindent Notice that confidence intervals do not directly correspond to the exact bounds used in our previous approximation algorithm. Still, we employ the same stopping criterion, that is, we run the Monte Carlo simulation until the width of the confidence interval is at most $\delta_p$.

A similar algorithm (without the use of confidence intervals) 
was also used in the context of biological networks (not represented as Prolog programs) by~\cite{Sevon06}. 
The use of a Monte Carlo method for probabilistic logic programs was suggested already by~\cite{Dantsin}, although he neither provides details nor reports on an implementation.
Our approach differs from the MCMC method for Stochastic Logic Programs (SLPs) introduced by~\cite{Cussens00}  in that we do not use a Markov chain, but restart from scratch for each sample. Furthermore, SLPs are different in that they directly define a distribution over all proofs of a query. Investigating similar probabilistic backtracking approaches for ProbLog is a promising future research direction.

\section{Implementation}\label{sec:implementation}
\begin{figure}[t]
\centering
\includegraphics[scale=0.6]{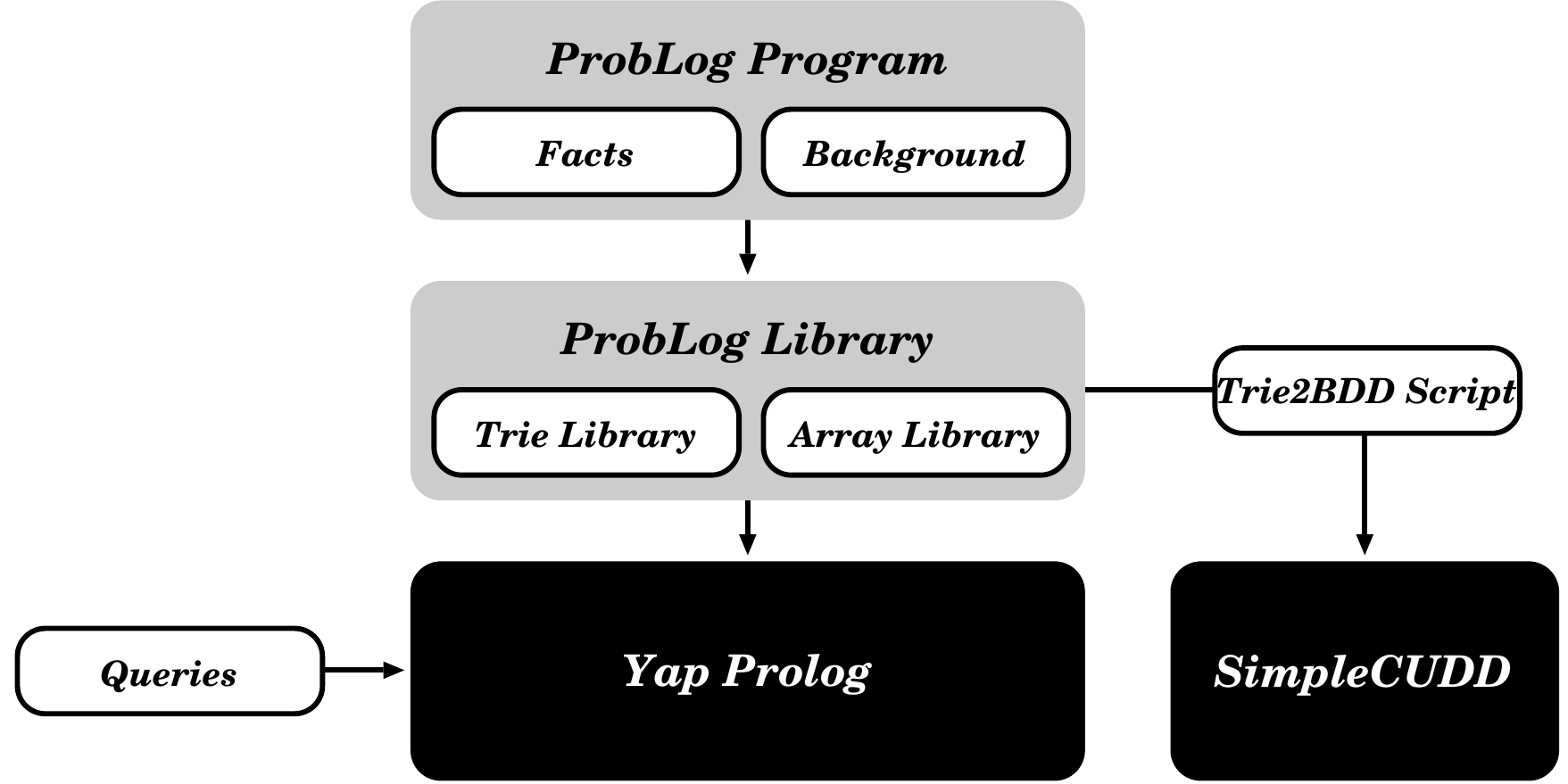}\label{fig:problog}
\caption{ProbLog Implementation: A ProbLog program (top) requires
  the ProbLog library which in turn relies on functionality from the
  tries and array libraries. ProbLog queries (bottom-left) are sent to
  the YAP engine, and may require calling the BDD library CUDD via SimpleCUDD.}
\label{fig:problog_imp}
\end{figure}

This section discusses the main building blocks used to implement ProbLog on top of the YAP-Prolog system. An overview is shown in Figure~\ref{fig:problog_imp}, with a typical ProbLog program, including ProbLog facts and background knowledge (BK), at the top.

The implementation requires ProbLog programs to use the
\texttt{problog} module. Each program consists of a set of labeled 
facts and of unlabeled \emph{background knowledge}, a generic
Prolog program. Labeled facts are preprocessed as described below.
Notice that the implementation requires all queries to non-ground probabilistic facts to be ground on calling.

In contrast to standard Prolog queries, where one is interested in
answer substitutions, in ProbLog one is primarily interested in a probability. As
discussed before, two common ProbLog queries ask for the most likely
explanation and its probability, and the probability of whether a
query would have an answer substitution. We have discussed two very
different approaches to the problem:
\begin{itemize}
\item In exact inference, $k$-best and bounded approximation, the engine explicitly reasons about
  probabilities of proofs. The challenge is how to compute the
  probability of each individual proof, store a large number of
  proofs, and compute the probability of sets of proofs.
\item In Monte Carlo, the probabilities of facts are used to sample from ProbLog programs. The
  challenge is how to compute a sample quickly, in a way that
  inference can be as efficient as possible.
\end{itemize}
ProbLog programs execute from a top-level query and are driven through a ProbLog query. The inference algorithms discussed above can be abstracted as follows:
\begin{itemize}
\item Initialise the inference algorithm;
 \item While probabilistic inference did not converge:
  \begin{itemize}
  \item initialise a new query;
  \item execute the query, instrumenting every ProbLog call in the current proof. Instrumentation is required for recording the ProbLog facts required by a proof, but may also be used by the inference algorithm to stop proofs (e.g.,\ if the current probability is lower than a bound);
  \item process success or exit substitution;
  \end{itemize}
\item Proceed to the next step of the algorithm: this may be trivial or may require calling an external solver, such as a BDD tool, to compute a probability.
\end{itemize}
Notice that the current ProbLog implementation relies on the Prolog
engine to efficiently execute goals. On the other hand, and in
contrast to most other probabilistic language implementations, in ProbLog there
is no clear separation between logical and probabilistic inference:
in a fashion similar to constraint logic programming, probabilistic
inference can drive logical inference.

From a Prolog implementation perspective, ProbLog poses a number of interesting challenges.  First, labeled facts have to be efficiently compiled to allow mutual calls between the Prolog program and the ProbLog engine.  Second, for exact inference, $k$-best and bounded approximation, sets of proofs have to be manipulated and transformed into BDDs.  Finally, Monte Carlo simulation requires representing and manipulating samples.  We discuss these issues next.

\subsection{Source-to-source transformation}

We use the \texttt{term\_expansion} mechanism to allow Prolog
calls to labeled facts, and for labeled facts to call the ProbLog
engine. As an example, the program:
\begin{equation}
\begin{array}{l}
\mathtt{0\ldotp715::edge('PubMed\_2196878','MIM\_609065')\ldotp}\\
\mathtt{0\ldotp659::edge('PubMed\_8764571','HGNC\_5014')\ldotp}\\
\end{array}
\end{equation}
\noindent
would be compiled as:
\begin{equation}
\begin{array}{lll}
  \mathtt{edge(A,B)} &\mathtt{:-} &  \mathtt{problog\_edge(ID,A,B,LogProb),}\\
  & & \mathtt{grounding\_id(edge(A,B),ID,GroundID),}\\
  & & \mathtt{add\_to\_proof(GroundID,LogProb)\ldotp}\\
  & & \\
  \multicolumn{3}{l}{\mathtt{problog\_edge(0,'PubMed\_2196878','MIM\_609065',-0\ldotp3348)\ldotp}}    \\
  \multicolumn{3}{l}{\mathtt{problog\_edge(1,'PubMed\_8764571','HGNC\_5014',-0\ldotp4166)\ldotp}}   \\
\end{array}
\end{equation}
\noindent
Thus, the internal representation of each fact contains an identifier, the original arguments, and the logarithm of the probability\footnote{We use the logarithm to avoid numerical problems when calculating  the probability of a derivation, which is used to drive inference.}. The \texttt{grounding\_id} procedure will create and store a grounding specific identifier for each new grounding of a non-ground probabilistic fact encountered during proving, and retrieve it on repeated use. For ground probabilistic facts, it simply returns the identifier itself.   The \texttt{add\_to\_proof} procedure updates the data structure representing the current path through the search space, i.e., a queue of identifiers ordered by first use, together with  its probability.  Compared to the original meta-interpreter based implementation of~\cite{DeRaedt07}, the main benefit of source-to-source transformation is better scalability, namely by having a compact representation of the facts for the YAP engine~\cite{DBLP:conf/padl/Costa07} and by allowing access to the YAP indexing mechanism~\cite{jit-index}.

\subsection{Proof Manipulation}

Manipulating proofs is critical in ProbLog.  We represent each proof
as a queue containing the identifier of each
different ground probabilistic fact used in the proof, ordered by
first use. The implementation requires calls to non-ground probabilistic facts to be ground, and during proving maintains a table of groundings used within the current query together with their identifiers. Grounding identifiers are based on the fact's identifier extended with a grounding number, i.e.~$5\_1$ and $5\_2$ would refer to different groundings of the non-ground fact with identifier $5$. 
In our implementation, the queue is stored in a backtrackable global variable, which is updated by calling \texttt{add\_to\_proof} with an identifier for the current ProbLog fact. We thus exploit Prolog's backtracking mechanism to avoid recomputation of shared proof prefixes when exploring the space of proofs. 
Storing a proof is simply a question of adding the value of the variable to a store.

As we have discussed above, the actual number of proofs can grow very quickly. ProbLog compactly represents a proof as a list of numbers.  We would further like to have a scalable implementation of \emph{sets} of proofs, such that we can compute the joint \emph{probability} of large sets of proofs efficiently. Our representation for sets of proofs and our algorithm for computing the probability of such a set are discussed next. 
 
\subsection{Sets of Proofs}

Storing and manipulating proofs is critical in ProbLog. 
When manipulating proofs, the key operation is often \emph{insertion}:
we would like to add a proof to an existing set of proofs. Some
algorithms, such as exact inference or Monte Carlo, only manipulate
complete proofs. Others, such as bounded approximation, require adding
partial derivations too. The nature of the SLD-tree means that proofs
tend to share both a prefix and a suffix. Partial proofs tend to share
prefixes only. This suggests using \emph{tries} to maintain the set of
proofs. We use the YAP implementation of tries for this task, based
itself on XSB Prolog's work on tries of terms~\cite{RamakrishnanIV-99}, which we briefly summarize here. 

Tries~\cite{Fredkin-62} were originally invented to index
dictionaries, and have since been generalised to index recursive data
structures such as terms. Please refer
to~\cite{Bachmair-93,Graf-96,RamakrishnanIV-99} for the use of tries in
automated theorem proving, term rewriting and tabled logic
programs. An essential property of the trie data structure is that
common prefixes are stored only once. A trie is a tree structure where
each different path through the trie data units, the \emph{trie
  nodes}, corresponds to a term described by the tokens labelling the
nodes traversed. For example, the tokenized form of the term
$f(g(a),1)$ is the sequence of 4 tokens: $f/2$, $g/1$, $a$ and
$1$. Two terms with common prefixes will branch off from each other at
the first distinguishing token.

Trie's internal nodes are four field data structures, storing  
the node's token, a pointer to the node's
first child, a pointer to the node's parent and a
pointer to the node's next sibling, respectively. Each
internal node's outgoing transitions may be determined by following
the child pointer to the first child node and, from there, continuing
sequentially through the list of sibling pointers. When a list of
sibling nodes becomes larger than a threshold value (8 in our
implementation), we dynamically index the nodes through a hash table
to provide direct node access and therefore optimise the
search. Further hash collisions are reduced by dynamically expanding
the hash tables. Inserting a term requires in the worst case
allocating as many nodes as necessary to represent its complete
path. On the other hand, inserting repeated terms requires traversing
the trie structure until reaching the corresponding leaf node, without
allocating any new node.

In order to minimize the number of nodes when storing proofs in a
trie, we use Prolog lists to represent proofs. For example, 
a ProbLog proof $[3, 5\_1, 7, 5\_2]$ uses ground
fact 3, a first grounding of fact 5, ground fact 7 and another
grounding of fact 5, that is, list elements in proofs are always
either integers or two integers with an underscore in between.

\begin{figure}[t]
\centering
\includegraphics[scale=0.55]{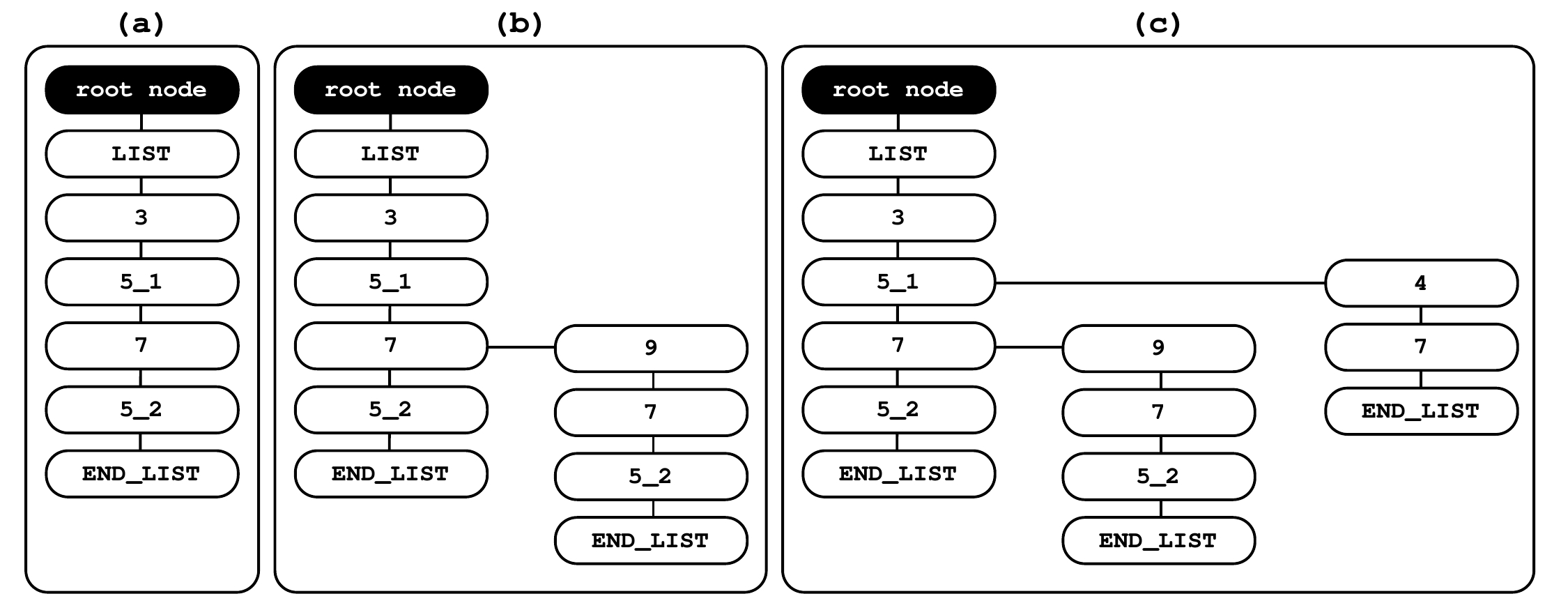}
\caption{Using tries to store proofs. Initially, the trie contains the
  root node only. Next, we store the proofs: 
  (a) $[3, 5\_1, 7, 5\_2]$;
  (b) $[3, 5\_1, 9, 7, 5\_2]$; and
  (c) $[3, 4, 7]$.}
\label{fig:trie_proofs}
\end{figure}

Figure~\ref{fig:trie_proofs} presents an example of a trie storing
three proofs. Initially, the trie contains the root node only. Next,
we store the proof $[3, 5\_1, 7, 5\_2]$ and six nodes (corresponding
to six tokens) are added to represent it
(Figure~\ref{fig:trie_proofs}(a)). The proof $[3, 5\_1, 9, 7, 5\_2]$
is then stored which requires seven nodes. As it shares a common
prefix with the previous proof, we save the three initial nodes common
to both representations (Figure~\ref{fig:trie_proofs}(b)). The proof
$[3, 4, 7]$ is stored next and we save again the two initial nodes
common to all proofs (Figure~\ref{fig:trie_proofs}(c)).

\subsection{Binary Decision Diagrams}
\label{sec:BDD}

\begin{figure}[t]
\centering
\includegraphics[scale=0.4]{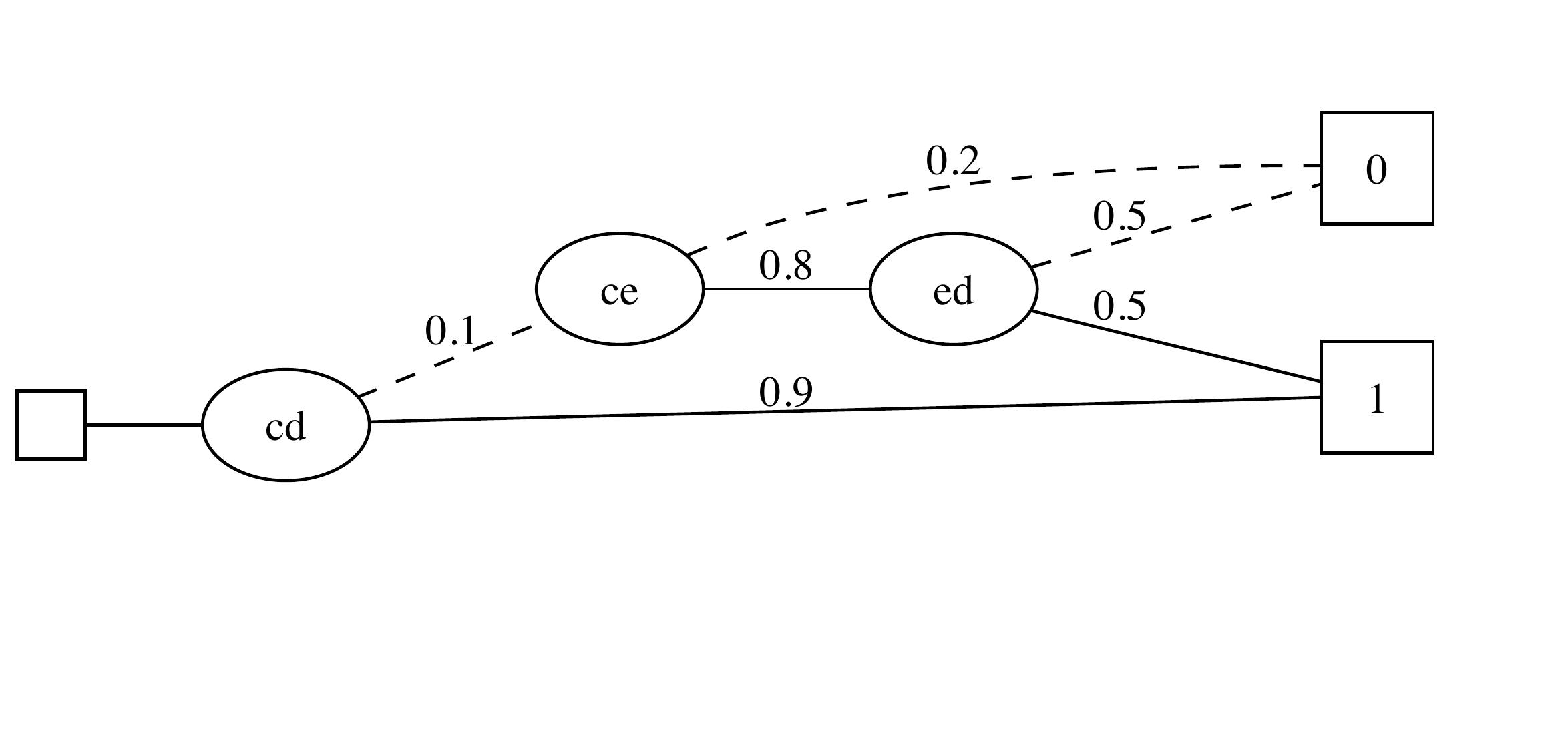}
\caption{Binary Decision Diagram encoding the DNF formula $cd \vee (ce \wedge
 ed)$, corresponding to the two proofs of query \emph{path(c,d)} in
 the example graph. An internal node labeled $xy$ represents the Boolean
 variable for the edge between $x$ and~$y$, solid/dashed edges
 correspond to values true/false and are labeled with the probability that the variable takes this value.}\label{fig:BDD}
\end{figure}

To efficiently compute the probability of a DNF formula representing a set of proofs, 
our implementation represents this formula as a reduced ordered Binary Decision Diagram (BDD)~\cite{Bryant86}, which  can be viewed as a compact encoding of a Boolean decision tree. 
Given a fixed variable ordering, a Boolean function $f$ can be represented as
a full Boolean decision tree, where each node on the $i$th level is
labeled with the $i$th variable and has two children called low and
high.  Leaves are labeled by the outcome of $f$ for the variable
assignment corresponding to the path to the leaf, where in each node
labeled $x$, the branch to the low (high) child is taken if variable
$x$ is assigned 0 (1).  Starting from such a tree, one obtains a BDD
by merging isomorphic subgraphs and deleting redundant nodes until no
further reduction is possible. A node is redundant if the subgraphs
rooted at its children are isomorphic.  Figure~\ref{fig:BDD} shows the
BDD for the existence of a path between \emph{c} and \emph{d} in our
earlier example.

We use  SimpleCUDD\footnote{\url{http://www.cs.kuleuven.be/~theo/tools/simplecudd.html}} as a wrapper tool for the 
BDD package
CUDD\footnote{\url{http://vlsi.colorado.edu/~fabio/CUDD}} to construct
and evaluate BDDs. More precisely, the trie representation of the DNF
is translated to a BDD generation script, which is processed by SimpleCUDD to build the 
BDD using CUDD primitives. It is executed via Prolog's shell utility, and results
are reported via shared files. 

\begin{figure}
\centering
\subfigure[]{\includegraphics[scale=0.25]{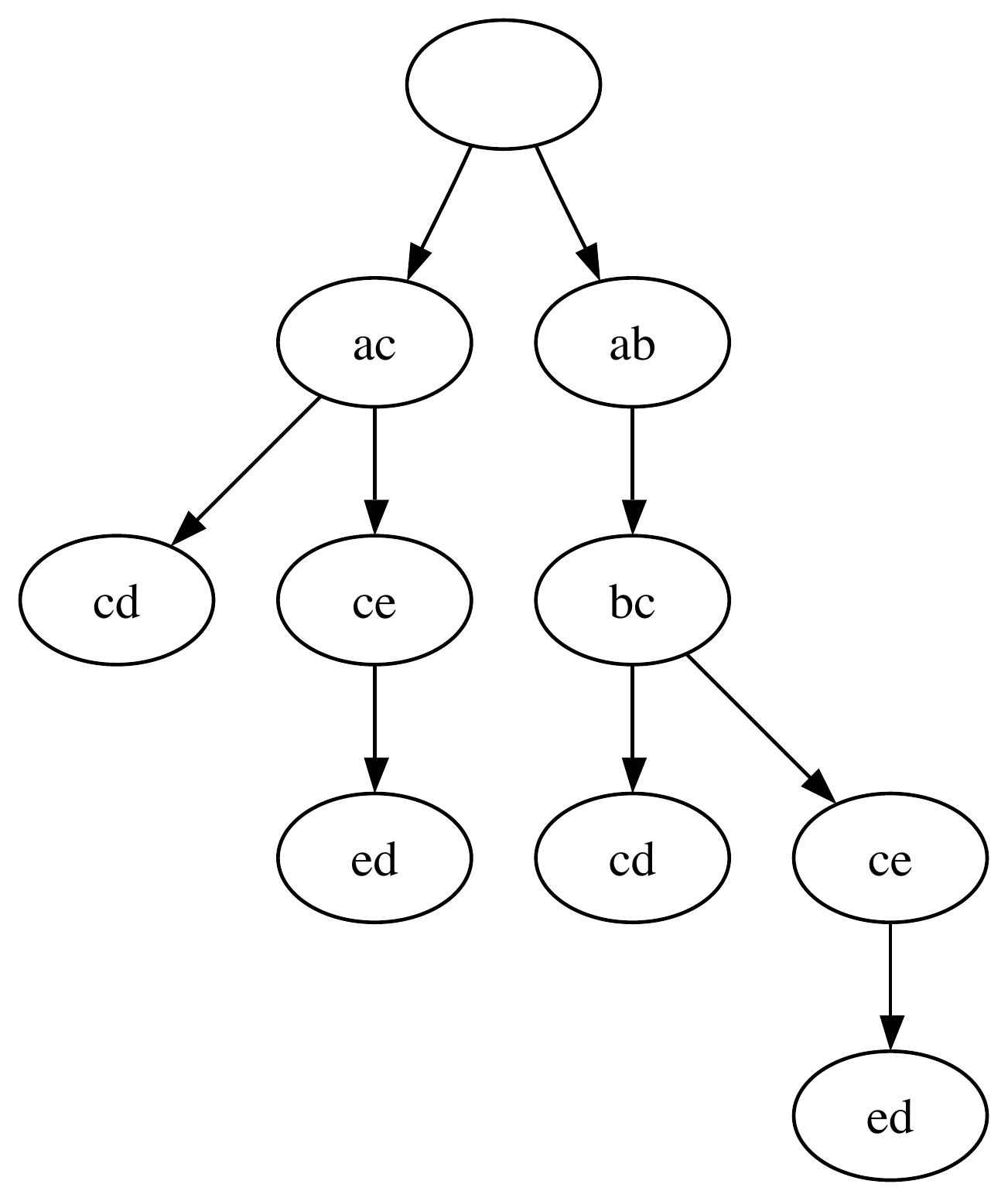}}
\subfigure[]{\includegraphics[scale=0.25]{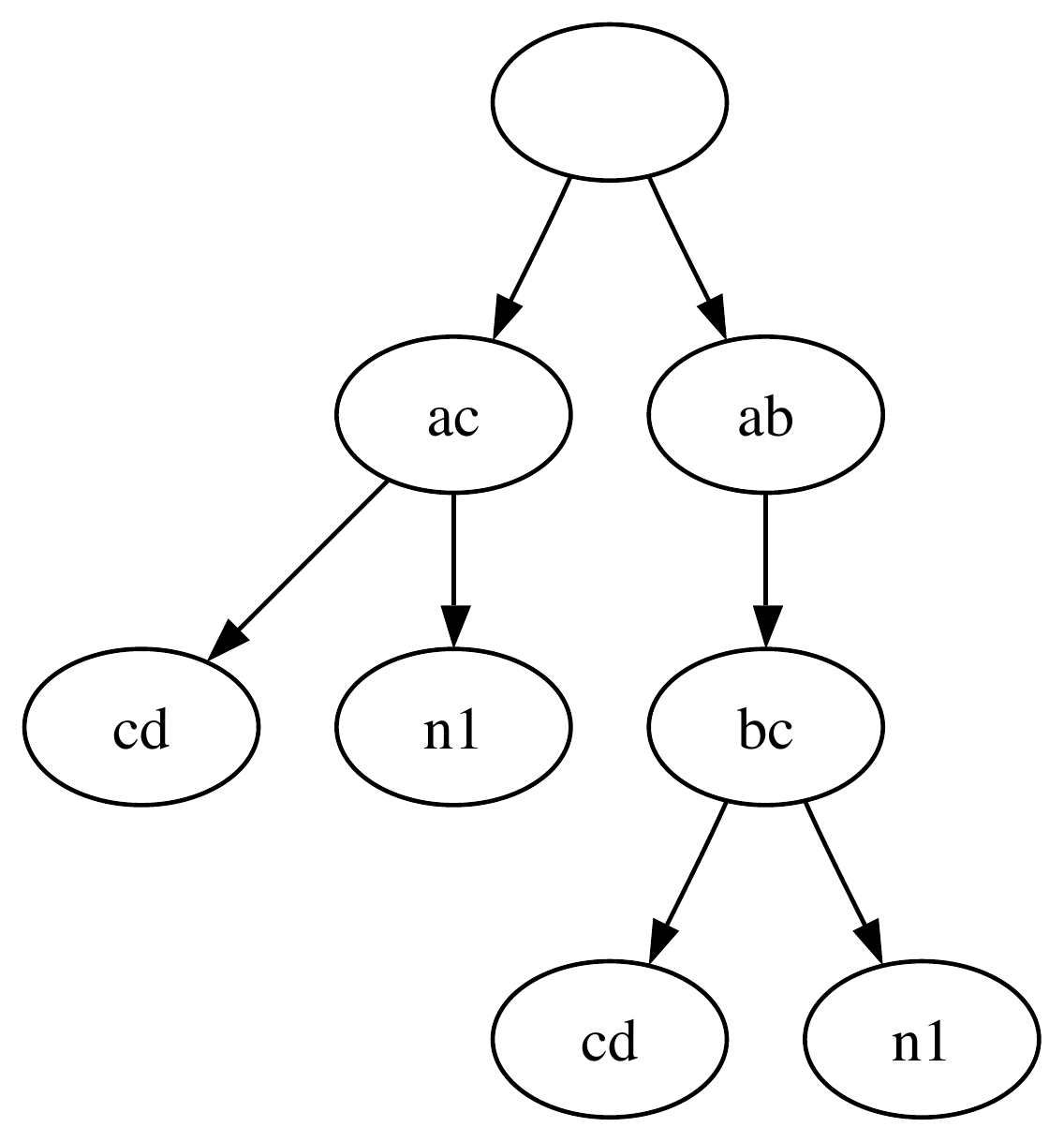}}
\subfigure[]{\includegraphics[scale=0.25]{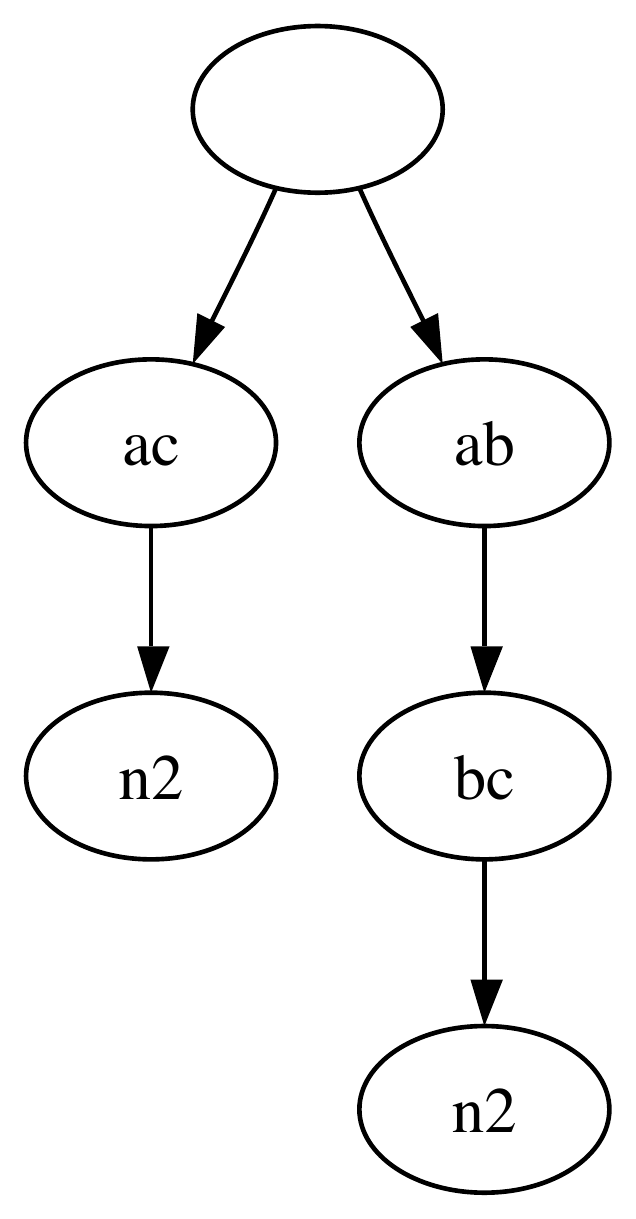}}
\subfigure[]{\includegraphics[scale=0.25]{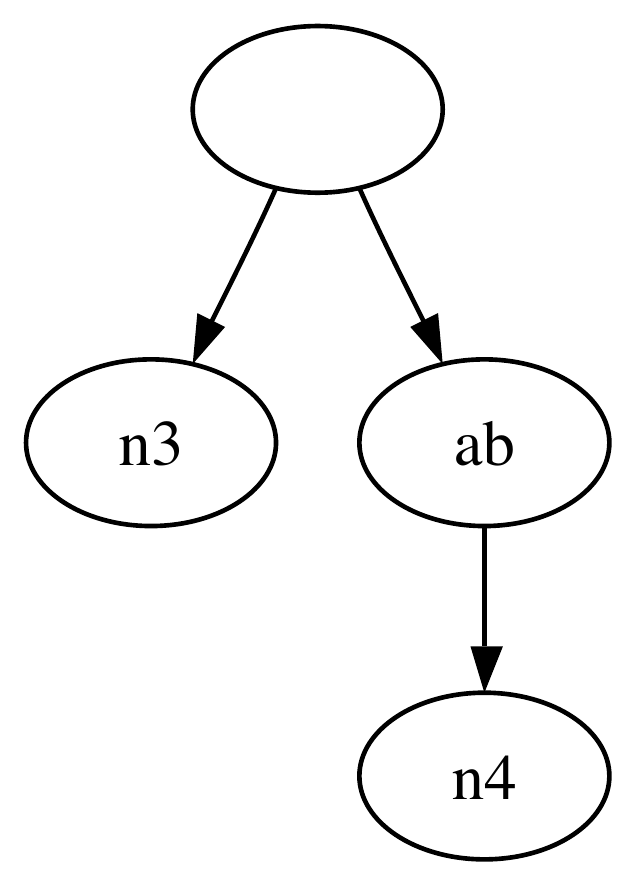}}
\subfigure[]{\includegraphics[scale=0.25]{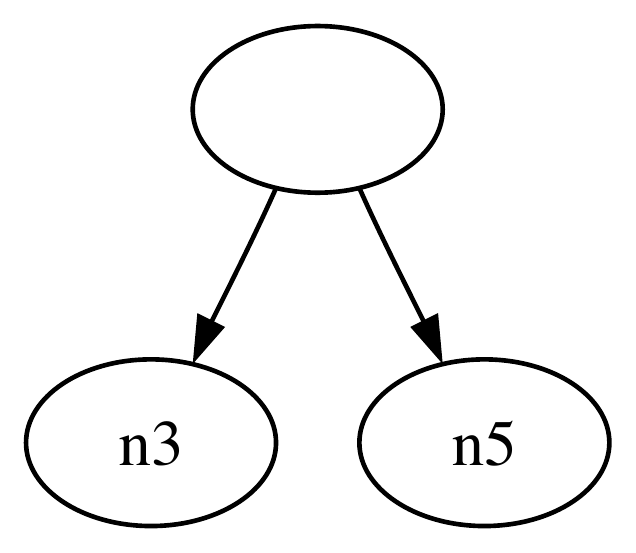}}
\caption{Translating the DNF for \emph{path(a,d)}.}
\label{fig:trie2bdd}
\end{figure}

\begin{algorithm}[t]
  \caption{Translating a trie $T$ representing a DNF to a BDD generation script. \textsc{Replace}$(T,C,n_i)$ replaces each occurence of $C$ in $T$ by $n_i$.}
\label{alg:trie2bdd}
\begin{algorithmic}
\FUNCTION{\textsc{Translate}(trie $T$)}
\STATE $i := 1$
\WHILE{$\neg leaf(T)$}
\STATE $S_{\wedge} := \{(C,P)|C $ leaf in $T$ and single child of its parent $P \}$
\FORALL{$(C,P)\in S_{\wedge}$}
\STATE write $n_i = P\wedge C$
\STATE $T := \textsc{Replace}(T,(C,P),n_i)$
\STATE $i := i + 1$
\ENDFOR
\STATE $S_{\vee} := \{[C_1,\ldots,C_n]|$ leaves $C_j $ are all the children of some parent $P$ in $T\}$
\FORALL{$[C_1,\ldots,C_n]\in S_{\vee}$}
\STATE write $n_i = C_1 \vee \ldots \vee C_n$
\STATE $T := \textsc{Replace}(T,[C_1,\ldots,C_n],n_i)$
\STATE $i := i + 1$
\ENDFOR
\ENDWHILE
\STATE write $top = n_{i-1}$
\ENDFUNCTION
\end{algorithmic}
\end{algorithm} 
During the generation of the code, it is crucial to exploit the
structure sharing (prefixes and suffixes) already in the trie
representation of a DNF formula, otherwise CUDD computation time
becomes extremely long or memory overflows quickly. 
Since CUDD builds BDDs by joining smaller BDDs using logical operations, the trie
is traversed bottom-up to successively generate code for all its
subtrees. Algorithm~\ref{alg:trie2bdd} gives the details of this procedure. Two types of operations are used to combine nodes. 
The first creates conjunctions of leaf nodes and their parent if the leaf is a single child, the second creates disjunctions of all child nodes of a node if these child nodes are all leaves. 
In both cases, a subtree that occurs multiple times in the trie is
translated only once, and the resulting BDD is used for all occurrences
of that subtree. Because of the optimizations in CUDD, the resulting
BDD can have a very different structure than the trie.
The translation for query \emph{path(a,d)} in our example graph  is illustrated  in Figure~\ref{fig:trie2bdd}, it results in the following script:
\begin{eqnarray*}
n1 & = & ce \wedge ed\\
n2 & = & cd \vee n1\\
n3 & = & ac \wedge n2\\
n4 & = & bc \wedge n2\\
n5 & = & ab \wedge n4\\
n6 & = & n3 \vee n5\\
top & = & n6
\end{eqnarray*}

After CUDD has generated the BDD, the probability of a formula is
calculated by traversing the BDD, in each node summing
the probability of the high and low child, weighted by the probability
of the node's variable being assigned true and false
respectively, cf.~Algorithm~\ref{alg:calcprob}. Intermediate results are cached, and the algorithm has a
time and space complexity linear in the size of the BDD.
\begin{algorithm}[t]
  \caption{Calculating the probability of a BDD.}
\label{alg:calcprob}
\begin{algorithmic}
\FUNCTION{\textsc{Probability}(BDD  node $n$ )}
\STATE  If $n$ is the 1-terminal then return 1
\STATE  If $n$ is the 0-terminal then return 0
\STATE  let $h$ and $l$ be the high and low children of $n$
\STATE  $prob(h) :=$ call \textsc{Probability}($h$)
\STATE  $prob(l) :=$ call \textsc{Probability}($l$) 
\STATE  return $p_n \cdot prob(h) + (1-p_n) \cdot prob(l)$ 
\ENDFUNCTION
\end{algorithmic}
\end{algorithm}
For illustration, consider again Figure~\ref{fig:BDD}. The algorithm starts by assigning probabilities $0$ and $1$ to the $0$-~and $1$-leaf respectively. The node labeled $ed$ has probability $0\ldotp5\cdot1+0\ldotp5\cdot0=0\ldotp5$, node $ce$ has probability $0\ldotp8\cdot0\ldotp5+0\ldotp2\cdot0=0\ldotp4$; finally, node $cd$, and thus the entire formula, has probability $0\ldotp9\cdot1+0\ldotp1\cdot0\ldotp4=0\ldotp94$.

\subsection{Monte Carlo}
The Monte Carlo implementation is shown in Algorithm~\ref{alg:mc}. It receives a query $q$, an acceptance threshold $\delta_p$ and a constant $m$ determining the number of samples generated per iteration. At the end of each iteration, it estimates the probability $p$ as the fraction of programs sampled over all previous iterations that entailed the query, and the confidence interval width to be used in the stopping criterion as explained in Section~\ref{sec:mc_method}. 
\begin{algorithm}[t]
  \caption{Monte Carlo Inference.}
\label{alg:mc}
\begin{algorithmic}
\FUNCTION{\textsc{MonteCarlo}(query $q$, interval width $\delta_p$, constant $m$)}
\STATE $c = 0$; $i = 0$; $p = 0$; $\delta = 1$; 
\WHILE{$\delta > \delta_p$}
\STATE Generate a sample $P'$;
\IF{$P'\models q$} 
\STATE $c:=c+1;$ 
\ENDIF
\STATE $i:=i+1$;
\IF{$i$ mod $m ==0$}  
\STATE $p := c/i$
\STATE $\delta := 2\times\sqrt{\frac{p\cdot(p-1)}{i}}$
\ENDIF
\ENDWHILE
\STATE return $p$
\ENDFUNCTION
\end{algorithmic}
\end{algorithm} 
Monte Carlo execution is quite different from the approaches discussed before, as the two main steps are \textbf{(a)} generating a sample program and \textbf{(b)} performing standard refutation on the sample. Thus, instead of combining large numbers of proofs, we need to manipulate large numbers of different programs or samples.

Our first approach was to generate a complete sample and to check for a proof. In order to accelerate the process, proofs were cached in a trie to skip inference on a new sample. If no proofs exist on a cache, we call the standard Prolog refutation procedure. Although this approach works rather well for small databases, it does not scale to larger databases where just generating  a new sample requires walking through millions of facts.

We observed that even in large programs proofs are often quite short, i.e., we only need to verify whether facts from a small fragment of the database are in the sample. This suggests that it may be a good idea to take advantage of the independence between facts and generate the sample \emph{lazily}: we verify whether a fact is in the sample only when we need it for a proof.  YAP represents samples compactly as a three-valued array with one field for each fact, where $0$ means the fact was not yet sampled, $1$ it was already sampled and belongs to the sample, $2$ it was already sampled and does not belong to the sample.
In this implementation:
\begin{enumerate}
\item New samples are generated by resetting the sampling array.
\item At every call to \texttt{add\_to\_proof}, given the current ProbLog literal $f$:
  \begin{enumerate}
  \item if $s[f] == 0 $, $s[f] = sample(f)$;
  \item if $s[f] == 1$, succeed;
  \item if $s[f] == 2$, fail;
  \end{enumerate}
\end{enumerate}
Note that as fact identifiers are used to access the array, the approach cannot directly be used for non-ground facts.
The current implementation of Monte Carlo therefore uses the internal database to store the result of sampling different groundings of such facts.

\section{Experiments}\label{sec:experiments}
We performed experiments with our implementation of ProbLog in the context of the biological network obtained from the Biomine project~\cite{Sevon06}.  
We used two subgraphs extracted around three genes known to be connected
to the Alzheimer disease (HGNC numbers 983, 620 and 582) as well as
the full network. The smaller graphs were obtained querying Biomine for best paths of length 2 (resulting in graph \textsc{Small}) or all paths of length 3 (resulting in graph \textsc{Medium}) starting at one of the three genes. \textsc{Small} contains 79 nodes and 144 edges, \textsc{Medium} 5220 nodes and 11532 edges.
We used \textsc{Small} for a first comparison of our
algorithms on a small scale network where success probabilities can be calculated exactly. 
Scalability was evaluated using both \textsc{Medium} and the entire 
\textsc{Biomine} network with roughly 1,000,000 nodes and
6,000,000 edges. In all experiments, we queried for the probability that
two of the gene nodes mentioned above are connected, that is, we used queries such as \texttt{path('HGNC\_983','HGNC\_620',Path)}. We used the following definition of an acyclic path in our background knowledge:
\begin{equation}
\begin{array}{lll}
\mathtt{path(X,Y,A)} & \mathtt{:-} & \mathtt{path(X,Y,[X],A)},\\
\mathtt{path(X,X,A,A)\ldotp} & &\\
\mathtt{path(X,Y,A,R)} & \mathtt{:-} &   \mathtt{X~\backslash ==~Y},  \\ & & \mathtt{edge(X,Z),}   \\ & &  \mathtt{absent(Z,A),}   \\ & & \mathtt{path(Z,Y,[Z|A],R)\ldotp}\\
\end{array}
\end{equation}
As list operations to check for the absence of a node get expensive for long paths, we consider an alternative definition for use in Monte Carlo. It provides cheaper testing by using the internal database of YAP to store nodes on the current path under key \texttt{visited}:
\begin{equation}
\begin{array}{lll}
\mathtt{memopath(X,Y,A)} & \mathtt{:-} & \mathtt{eraseall(visited)},  \\ && \mathtt{memopath(X,Y,[X],A)\ldotp}\\
\mathtt{memopath(X,X,A,A)\ldotp} & &\\
\mathtt{memopath(X,Y,A,R)} & \mathtt{:-} & \mathtt{X~\backslash ==~Y}, \\ & & \mathtt{edge(X,Z),}    \\ & &  \mathtt{recordzifnot(visited,Z,\_),}\\
 & &    \mathtt{memopath(Z,Y,[Z|A],R)\ldotp}\\
\end{array}
\end{equation}
Finally, to assess performance on the full network for queries with smaller probabilities, we use the following definition of paths with limited length:
\begin{equation}
\begin{array}{lll}
\mathtt{lenpath(N,X,Y,Path)} & \mathtt{ :-} & 	\mathtt{lenpath(N,X,Y,[X],Path)\ldotp}\\
\mathtt{lenpath(N,X,X,A,A) } & \mathtt{ :-} & \mathtt{	N >= 0\ldotp}\\
\mathtt{lenpath(N,X,Y,A,P) } & \mathtt{ :-} & \mathtt{	X \backslash == Y},\\
&&	\mathtt{	N > 0},\\
&&	\mathtt{	edge(X,Z)},\\
&&	\mathtt{	absent(Z,A)},\\
&&	\mathtt{	NN\  is\  N-1},\\
&&	\mathtt{	lenpath(NN,Z,Y,[Z|A],P)\ldotp}
\end{array}
\end{equation}

All experiments were performed on a Core 2 Duo 2.4 GHz 4 GB machine running Linux. All times reported are in \texttt{msec} and do not include the time to load the graph into Prolog. The latter takes 20, 200 and 78140 \texttt{msec} for \textsc{Small}, \textsc{Medium} and \textsc{Biomine} respectively. Furthermore, as YAP indexes the database at query time, we query for the explanation probability of  \texttt{path('HGNC\_620','HGNC\_582',Path)} before starting runtime measurements. This takes 0, 50 and 25900 \texttt{msec} for \textsc{Small}, \textsc{Medium} and \textsc{Biomine} respectively.
We report $T_P$, the time spent  by
ProbLog to search for proofs, as well as $T_B$, the time spent to execute BDD programs (whenever meaningful). We also report the
estimated probability $P$. For approximate inference using  bounds, we
report exact intervals for $P$, and also include the number $n$ of
BDDs constructed. We set both the initial threshold and the shrinking
factor to $0\ldotp5$.
We computed $k$-probability for $k=1,2,\ldots,1024$. In the bounding algorithms, the error interval ranged 
between 10\% and 1\%. Monte Carlo recalculates confidence intervals  after  $m=1000$ samples. We also report the number $S$ of samples used.

\paragraph{Small Sized Sample}
\begin{table}[t]
\begin{tabular}{|c|rrr|rrr|rrr|}
\hline
path &
\multicolumn{3}{c|}{$983-620$} &
\multicolumn{3}{c|}{$983-582$} &
\multicolumn{3}{c|}{$620-582$} \\
{\bf k}&
\multicolumn{1}{c}{$T_P$} & \multicolumn{1}{c}{$T_B$} & \multicolumn{1}{c|}{$P$} &
\multicolumn{1}{c}{$T_P$} & \multicolumn{1}{c}{$T_B$} & \multicolumn{1}{c|}{$P$} &
\multicolumn{1}{c}{$T_P$} & \multicolumn{1}{c}{$T_B$} & \multicolumn{1}{c|}{$P$} \\
\hline
   1 &    0 &  13 & 0.07 &   0 &   7 & 0.05 &   0 &  26 & 0.66\\
   2 &    0 &  12 & 0.08 &   0 &   6 & 0.05 &   0 &   6 & 0.66\\
   4 &    0 &  12 & 0.10 &  10 &   6 & 0.06 &   0 &   6 & 0.86\\
   8 &   10 &  12 & 0.11 &   0 &   6 & 0.06 &   0 &   6 & 0.92\\
  16 &    0 &  12 & 0.11 &  10 &   6 & 0.06 &   0 &   6 & 0.92\\
  32 &   20 &  34 & 0.11 &  10 &  17 & 0.07 &   0 &   7 & 0.96\\
  64 &   20 &  74 & 0.11 &  10 &  46 & 0.09 &  10 &  38 & 0.99\\
 128 &   50 & 121 & 0.11 &  40 & 161 & 0.10 &  20 & 257 & 1.00\\
 256 &  140 & 104 & 0.11 &  80 & 215 & 0.10 &  90 & 246 & 1.00\\
 512 &  450 & 118 & 0.11 & 370 & 455 & 0.11 & 230 & 345 & 1.00\\
1024 & 1310 & 537 & 0.11 & 950 & 494 & 0.11 & 920 & 237 & 1.00\\\hline
\textbf{exact} & 670 & 450 & 0.11 & 8060 & 659 & 0.11 & 630 & 721 & 1.00\\\hline
\end{tabular}
\caption{$k$-probability on \textsc{Small}. } 
\label{tab:1}
\end{table}
We first compared our algorithms on \textsc{Small}. 
Table~\ref{tab:1} shows the results for $k$-probability and exact inference. Note
that nodes 620 and 582 are close to each other, whereas node 983 is farther apart. Therefore, connections involving the latter are less likely.
In this graph, we obtained good approximations using a small fraction of proofs (the queries have 13136, 155695 and 16048    proofs respectively).
 Our results also show a significant
increase in running times as ProbLog explores more paths in the graph,
both within the Prolog code and within the BDD code. The BDD running
times can vary widely, we may actually have large running times for
smaller BDDs, depending on BDD structure. However, using SimpleCUDD instead of the C++ interface used in~\cite{Kimmig08} typically decreases BDD time by at least one or two orders of magnitude.

Table~\ref{tab:2} gives corresponding results for bounded approximation. The
algorithm converges  quickly, as few proofs are needed and BDDs remain small. Note however that exact inference is competitive for this problem size. Moreover, we observe large speedups compared to the implementation with meta-interpreters used in~\cite{DeRaedt07}, where total runtimes to reach $\delta=0\ldotp01$ for these queries were 46234, 206400 and 307966 \texttt{msec} respectively.
Table~\ref{tab:3} shows the performance of the Monte Carlo estimator. On \textsc{Small}, Monte Carlo is the fastest approach. Already within the first 1000 samples a good approximation is obtained.

The experiments on \textsc{Small} thus confirm that the implementation on top of YAP-Prolog enables efficient probabilistic inference on small sized graphs.
\begin{table}[t]
\begin{tabular}{|c|rr|rr|rr|}
\hline     
path &
\multicolumn{2}{c|}{$983-620$} &
\multicolumn{2}{c|}{$983-582$} &
\multicolumn{2}{c|}{$620-582$} \\
$\delta$ &
\multicolumn{1}{c}{$T_P$~$T_B$~n} & \multicolumn{1}{c|}{$P$} &
\multicolumn{1}{c}{$T_P$~~~$T_B$~~~n} & \multicolumn{1}{c|}{$P$} &
\multicolumn{1}{c}{$T_P$~~$T_B$~~n} & \multicolumn{1}{c|}{$P$} \\
\hline
0.10 & 0~~48~~4 & [0.07,0.12] & 10~~~~~74~~~~6 & [0.06,0.11] & 0~~~~~25~~2 & [0.91,1.00] \\
0.05 & 0~~71~~6 & [0.07,0.11] &  0~~~~~75~~~~6 & [0.06,0.11] & 0~~~486~~4 & [0.98,1.00] \\
0.01 & 0~~83~~7 & [0.11,0.11] &  140~~3364~~10 & [0.10,0.11] & 60~~1886~~6 & [1.00,1.00] \\\hline
\end{tabular}
\caption{Inference using bounds on \textsc{Small}.  }
\label{tab:2}
\end{table}
\begin{table}[t]
\begin{tabular}{|c|rrr|rrr|rrr|}
\hline     
path &
\multicolumn{3}{c|}{$983-620$} &
\multicolumn{3}{c|}{$983-582$} &
\multicolumn{3}{c|}{$620-582$} \\
$\delta$ &
\multicolumn{1}{c}{$S$} & \multicolumn{1}{c}{$T_P$} & \multicolumn{1}{c|}{$P$} &
\multicolumn{1}{c}{$S$} & \multicolumn{1}{c}{$T_P$} & \multicolumn{1}{c|}{$P$} &
\multicolumn{1}{c}{$S$} & \multicolumn{1}{c}{$T_P$} & \multicolumn{1}{c|}{$P$} \\
\hline
0.10 &  1000 &  10 & 0.11 &  1000 &  10 & 0.11 & 1000 & 30 & 1.00\\
0.05 &  1000 &  10 & 0.11 &  1000 &  10 & 0.10 & 1000 & 20 & 1.00\\
0.01 & 16000 & 130 & 0.11 & 16000 & 170 & 0.11 & 1000 & 30 & 1.00\\\hline
\end{tabular}
\caption{Monte Carlo Inference on \textsc{Small}.  }
\label{tab:3}
\end{table}

\paragraph{Medium Sized Sample}
For graph \textsc{Medium} with around 11000 edges, exact inference is no longer feasible. Table~\ref{tab:1a}
again shows results for the $k$-probability.  Comparing these
results with the corresponding values from Table~\ref{tab:1}, we
observe that the estimated probability is higher now: this is natural,
as the graph has both more nodes and is more connected, therefore leading to many more possible explanations. This also explains the increase in running times.   Approximate
inference using bounds only reached loose bounds (with differences $>0\ldotp 2$) on queries involving node \texttt{'HGNC\_983'}, as upper bound formulae with more than 10 million conjunctions were encountered, which could not be processed.

The Monte Carlo estimator using the standard definition of \texttt{path/3} on \textsc{Medium} did not complete the first $1000$ samples within one hour.  A detailed analysis shows
that this is caused by some queries backtracking too heavily. Table~\ref{tab:3a} therefore reports results using the memorising version \texttt{memopath/3}. With this improved definition, Monte Carlo performs well: it obtains a good approximation in a few seconds. Requiring tighter bounds however can increase runtimes significantly.
\begin{table}[t]
\begin{tabular}{|c|rrr|rrr|rrr|}
\hline     
path &
\multicolumn{3}{c|}{$983-620$} &
\multicolumn{3}{c|}{$983-582$} &
\multicolumn{3}{c|}{$620-582$} \\
{\bf k} &
\multicolumn{1}{c}{$T_P$} & \multicolumn{1}{c}{$T_B$} & \multicolumn{1}{c|}{$P$} &
\multicolumn{1}{c}{$T_P$} & \multicolumn{1}{c}{$T_B$} & \multicolumn{1}{c|}{$P$} &
\multicolumn{1}{c}{$T_P$} & \multicolumn{1}{c}{$T_B$} & \multicolumn{1}{c|}{$P$} \\
\hline
   1 &  180 &  6 & 0.33 & 1620 & 6 & 0.30 &   10 &    6 & 0.92 \\
   2 &  180 &  6 & 0.33 & 1620 & 6 & 0.30 &   20 &    6 & 0.92 \\
   4 &  180 &  6 & 0.33 & 1630 & 6 & 0.30 &   10 &    6 & 0.92 \\
   8 &  220 &  6 & 0.33 & 1630 & 6 & 0.30 &   20 &    6 & 0.92 \\
  16 &  260 &  6 & 0.33 & 1660 & 6 & 0.30 &   30 &    6 & 0.99 \\
  32 &  710 &  6 & 0.40 & 1710 & 7 & 0.30 &  110 &    6 & 1.00 \\
  64 & 1540 &  7 & 0.42 & 1910 & 6 & 0.30 &  200 &    6 & 1.00 \\
 128 & 1680 &  6 & 0.42 & 2230 & 6 & 0.30 &  240 &    9 & 1.00 \\
 256 & 2190 &  7 & 0.55 & 2720 & 6 & 0.49 &  290 &  196 & 1.00 \\
 512 & 2650 &  7 & 0.64 & 3730 & 7 & 0.53 & 1310 &  327 & 1.00 \\
1024 & 8100 & 41 & 0.70 & 5080 & 8 & 0.56 & 3070 & 1357 & 1.00 \\
\hline
\end{tabular}
\caption{$k$-probability on \textsc{Medium}. }
\label{tab:1a}
\end{table}
\begin{table}[t]
\begin{tabular}{|c|rrr|rrr|rrr|}
\hline
memo &
\multicolumn{3}{c|}{$983-620$} &
\multicolumn{3}{c|}{$983-582$} &
\multicolumn{3}{c|}{$620-582$} \\
$\delta$ &
\multicolumn{1}{c}{$S$} & \multicolumn{1}{c}{$T_P$} & \multicolumn{1}{c|}{$P$} &
\multicolumn{1}{c}{$S$} & \multicolumn{1}{c}{$T_P$} & \multicolumn{1}{c|}{$P$} &
\multicolumn{1}{c}{$S$} & \multicolumn{1}{c}{$T_P$} & \multicolumn{1}{c|}{$P$} \\
\hline
0.10 &  1000 &  1180 & 0.78 &  1000 &  2130 & 0.76 & 1000 & 1640 & 1.00\\
0.05 &  2000 &  2320 & 0.77 &  2000 &  4230 & 0.74 & 1000 & 1640 & 1.00\\
0.01 & 29000 & 33220 & 0.77 & 29000 & 61140 & 0.77 & 1000 & 1670 & 1.00\\\hline
\end{tabular}
\caption{Monte Carlo Inference using \texttt{memopath/3} on \textsc{Medium}. }
\label{tab:3a}
\end{table}

\paragraph{Biomine Database}
The Biomine Database covers hundreds of thousands of entities and
millions of links. 
On \textsc{Biomine}, we therefore restricted our experiments to the approximations given by
$k$-probability and Monte Carlo. Given the results on \textsc{Medium}, we directly used \texttt{memopath/3} for Monte Carlo. Tables~\ref{tab:1c} and~\ref{tab:3b} show the results on the large network.
We observe that on this large graph, the number of possible paths is tremendous, which implies success probabilities practically equal to 1. Still, we observe that 
ProbLog's  branch-and-bound search to find the best solutions performs reasonably also on this size of network. However, runtimes for obtaining tight confidence intervals with Monte Carlo explode quickly even with the improved path definition.
\begin{table}[t]
\begin{tabular}{|c|rrr|rrr|rrr|}
\hline     
path &
\multicolumn{3}{c|}{$983-620$} &
\multicolumn{3}{c|}{$983-582$} &
\multicolumn{3}{c|}{$620-582$} \\
{\bf k} &
\multicolumn{1}{c}{$T_P$} & \multicolumn{1}{c}{$T_B$} & \multicolumn{1}{c|}{$P$} &
\multicolumn{1}{c}{$T_P$} & \multicolumn{1}{c}{$T_B$} & \multicolumn{1}{c|}{$P$} &
\multicolumn{1}{c}{$T_P$} & \multicolumn{1}{c}{$T_B$} & \multicolumn{1}{c|}{$P$} \\
\hline
   1 &   5,760 &     49 & 0.16 &     8,910 &      48 & 0.11 &      10 & 48 & 0.59 \\
   2 &   5,800 &     48 & 0.16 &    10,340 &      48 & 0.17 &     180 & 48 & 0.63 \\
   4 &   6,200 &     48 & 0.16 &    13,640 &      48 & 0.28 &     360 &  48 & 0.65 \\
   8 &   7,480 &     48 & 0.16 &    15,550 &      49 & 0.38 &     500 &  48 & 0.66 \\
  16 &  11,470 &     49 & 0.50 &    58,050 &      49 & 0.53 &     630 &    48 & 0.92 \\
  32 &  15,100 &     49 & 0.57 &   106,300 &      49 & 0.56 &   2,220 &   167 & 0.95 \\
  64 &  53,760 &     84 & 0.80 &   146,380 &     101 & 0.65 &   3,690 &   167 & 0.95 \\
 128 &  71,560 &    126 & 0.88 &   230,290 &     354 & 0.76 &   7,360 &   369 & 0.98 \\
 256 & 138,300 &    277 & 0.95 &   336,410 &     520 & 0.85 &  13,520 &  1,106 & 1.00 \\
 512 & 242,210 &    730 & 0.98 &   501,870 &   2,744 & 0.88 &  23,910 &  3,444 & 1.00 \\
1024 & 364,490 & 10,597 & 0.99 & 1,809,680 & 100,468 & 0.93 & 146,890 & 10,675 & 1.00 \\
\hline
\end{tabular}
\caption{$k$-probability on \textsc{Biomine}. }
\label{tab:1c}
\end{table}
\begin{table}[t]
\begin{tabular}{|c|rrr|rrr|rrr|}
\hline
memo &
\multicolumn{3}{c|}{$983-620$} &
\multicolumn{3}{c|}{$983-582$} &
\multicolumn{3}{c|}{$620-582$} \\
$\delta$ &
\multicolumn{1}{c}{$S$} & \multicolumn{1}{c}{$T_P$} & \multicolumn{1}{c|}{$P$} &
\multicolumn{1}{c}{$S$} & \multicolumn{1}{c}{$T_P$} & \multicolumn{1}{c|}{$P$} &
\multicolumn{1}{c}{$S$} & \multicolumn{1}{c}{$T_P$} & \multicolumn{1}{c|}{$P$} \\
\hline
0.10 & 1000 & 100,700 & 1.00 & 1000 & 1,656,660 & 1.00 & 1000 & 1,696,420 & 1.00\\
0.05 & 1000 & 100,230 & 1.00 & 1000 & 1,671,880 & 1.00 & 1000 & 1,690,830 & 1.00\\
0.01 & 1000 &  93,120 & 1.00 & 1000 & 1,710,200 & 1.00 & 1000 & 1,637,320 & 1.00\\\hline
\end{tabular}
\caption{Monte Carlo Inference using \texttt{memopath/3} on \textsc{Biomine}. }
\label{tab:3b}
\end{table}
Given that sampling a program that does not entail the query is extremely unlikely for the setting considered so far, we performed an additional experiment on \textsc{Biomine}, where we restrict the number of edges on the path connecting two nodes to a maximum of 2 or 3. Results are reported in Table~\ref{tab:shortpath}. As none of the resulting queries have more than 50 proofs, exact inference is much faster than Monte Carlo, which needs a higher number of samples to reliably estimate probabilities that are not close to $1$. 
\begin{table}[t]
\begin{tabular}{|c|rrr|rrr|rrr|}
\hline
len &
\multicolumn{3}{c|}{$983-620$} &
\multicolumn{3}{c|}{$983-582$} &
\multicolumn{3}{c|}{$620-582$} \\
$\delta$ &
\multicolumn{1}{c}{$S$} & \multicolumn{1}{c}{$T$} & \multicolumn{1}{c|}{$P$} &
\multicolumn{1}{c}{$S$} & \multicolumn{1}{c}{$T$} & \multicolumn{1}{c|}{$P$} &
\multicolumn{1}{c}{$S$} & \multicolumn{1}{c}{$T$} & \multicolumn{1}{c|}{$P$} \\
\hline
0.10 & 1000 &  21,400 & 0.04 &  1000 &  18,720 & 0.11 &  1000 &  19,150 & 0.58\\
0.05 & 1000 &  19,770 & 0.05 &  1000 &  20,980 & 0.10 &  2000 &  35,100 & 0.55\\
0.01 & 6000 & 112,740 & 0.04 & 16000 & 307,520 & 0.11 & 40000 & 764,700 & 0.55\\\hline
exact & - & 477 & 0.04 & - & 456 & 0.11 & - & 581 & 0.55 \\\hline \hline
0.10 &  1000 &  106,730 & 0.14 &  1000 &   105,350 & 0.33 & 1000 &  45,400 & 0.96\\
0.05 &  1000 &  107,920 & 0.14 &  2000 &   198,930 & 0.34 & 1000 &  49,950 & 0.96\\
0.01 & 19000 &2,065,030 & 0.14 & 37000 & 3,828,520 & 0.35 & 6000 & 282,400 & 0.96\\\hline
exact & - & 9,413 & 0.14 & - & 9,485 & 0.35 & - & 15,806 & 0.96\\\hline
\end{tabular}
\caption{Monte Carlo inference for different values of $\delta$ and exact inference using \texttt{lenpath/4} with length at most $2$ (top) or $3$ (bottom) on \textsc{Biomine}. For exact inference, runtimes include both Prolog and BDD time.}
\label{tab:shortpath}
\end{table}

Altogether, the experiments confirm that our implementation provides efficient inference algorithms for ProbLog that scale to large databases. Furthermore, compared to the original implementation of~\cite{DeRaedt07}, we obtain large speedups in both the Prolog and the BDD part, thereby opening new perspectives for applications of ProbLog. 

\section{Conclusions}\label{sec:conclusion}
ProbLog is a simple but elegant probabilistic logic programming  
language that allows one to explicitly represent uncertainty by means of probabilistic facts denoting independent random variables.  The language  
is a simple and natural extension of the logic programming  
language Prolog.
We presented an efficient
implementation of the ProbLog language on top of the YAP-Prolog system
that is designed to scale to large sized problems.   We showed that
ProbLog can  be used to obtain both explanation and (approximations  
of) success
probabilities for queries on a large database. To the best of our
knowledge, ProbLog is the first example of a probabilistic logic
programming system that can execute queries on such large databases.
Due to the use of BDDs for addressing the disjoint-sum-problem, the initial implementation of ProbLog used in~\cite{DeRaedt07} already  
scaled up much better than alternative implementations such as Fuhr's  
pD engine HySpirit~\cite{Fuhr00}.
The tight integration in YAP-Prolog presented here leads to further speedups  
in runtime of several orders of magnitude.

Although we focused on connectivity queries and Biomine in this
work, similar problems are found across many domains; we
believe that the techniques presented   apply to a wide
variety of queries and databases because
ProbLog provides a clean separation between background knowledge and
what is specific to the engine. As shown for Monte Carlo inference,
such an interface can be very useful to improve performance
as it allows incremental refinement of background knowledge,
e.g., graph procedures. Initial experiments with Dijkstra's algorithm  
for finding the explanation probability are very promising.

  ProbLog is closely related to some alternative formalisms such as  
PHA and ICL~\cite{Poole:93,Poole00}, pD~\cite{Fuhr00}
and  PRISM~\cite{SatoKameya:01}  as their semantics are all based on  
Sato's distribution semantics  even though there exist also some  
subtle differences.
  However,  ProbLog is  -- to the best of the authors' knowledge --  
the first implementation that tightly integrates Sato's  original  
distribution semantics~\cite{Sato:95}  in a state-of-the-art Prolog  
system without making additional restrictions (such as the exclusive  
explanation assumption made in PHA and PRISM).  
As ProbLog, both PRISM and the ICL implementation AILog2 use a  two-step approach to inference, where proofs are collected in the first phase, and probabilities are calculated once all proofs are known. AILog2 is a meta-interpreter implemented in SWI-Prolog for didactical purposes, where the disjoint-sum-problem is tackled using a symbolic disjoining technique~\cite{Poole00}. PRISM, built on top of B-Prolog, requires programs to be written such that alternative explanations for queries are mutually exclusive. PRISM uses a meta-interpreter to collect proofs in a hierarchical datastructure called explanation graph. As proofs are mutually exclusive, the explanation graph directly mirrors the sum-of-products structure of probability calculation~\cite{SatoKameya:01}.  
ProbLog is the first probabilistic logic programming system using BDDs as a basic datastructure for probability calculation, a principle that  
receives increased interest in the probabilistic logic learning community, cf.~for instance~\cite{Riguzzi,sato:ilp08}.

Furthermore,
as compared to  SLPs~\cite{Muggleton96}, CLP($\cal BN$) \cite{Costa03:uai}, and BLPs~\cite{Kersting08}, ProbLog is a much  
simpler and in a sense more primitive probabilistic programming   
language. Therefore, the relationship between probabilistic logic  
programming and ProbLog is, in a sense, analogous to that between
logic programming and Prolog. From this perspective, it is our hope  
and goal to further develop ProbLog so that it can be used as a  
general purpose programming language with an efficient implementation  
for use in statistical relational learning~\cite{Getoor07} and  
probabilistic  programming~\cite{DeRaedt-PILPbook}.
One important use of such a probabilistic programming language is as a  
target language in which other formalisms can be efficiently compiled.
For instance, it has already been shown that  CP-logic~\cite{Vennekens}, a recent elegant probabilistic knowledge  
representation language based on a probabilistic extension of  clausal  
logic, can be compiled into ProbLog~\cite{Riguzzi} and it is well-known that SLPs~\cite{Muggleton96} can be compiled into Sato's PRISM,
which is closely related to ProbLog. Further evidence is provided  
in~\cite{DeRaedt-NIPSWS08}.

Another, related use of ProbLog is as a vehicle for developing  
learning and mining algorithms and tools~\cite{Kimmig07,DeRaedt08MLJ,Gutmann08,Kimmig09,DeRaedt-IQTechReport}.
In the context of probabilistic representations~\cite{Getoor07,DeRaedt-PILPbook}, one typically distinguishes  
two types of learning: parameter estimation
and structure learning. In parameter estimation in the context of  
ProbLog and PRISM, one starts from a set of queries and the logical  
part of the program and the problem is to find good  estimates of the  
parameter values, that is, the probabilities of the probabilistic  
facts in the program. \cite{Gutmann08} introduces a gradient descent approach to parameter learning for ProbLog that extends the BDD-based methods discussed here. 
In structure learning, one also starts from  
queries but has to find the logical part of the program as well.  
Structure learning is therefore closely related to inductive logic  
programming.
The limiting factor in statistical relational learning and  
probabilistic logic learning is often the efficiency of inference, as learning requires
repeated computation of the probabilities of many queries.
Therefore, improvements on inference in probabilistic programming  
implementations have an immediate effect on learning.
The above compilation approach also raises the interesting and largely  
open question whether
not only inference problems for alternative formalisms can be compiled  
into ProbLog but whether it  is also possible to compile learning  
problems for these logics
into learning problems for ProbLog.

  Finally, as ProbLog, unlike
PRISM and PHA, deals with the disjoint-sum-problem, it is interesting
to study how program transformation and analysis techniques could be
used to optimize ProbLog programs, by detecting and taking into
account situations where some conjunctions are disjoint.  At the same  
time, we currently investigate how 
tabling, one of the keys to PRISM's efficiency, can be incorporated in ProbLog~\cite{Mantadelis09,Kimmig-SRL09}.

\subsubsection*{Acknowledgements} 
We would like to thank Hannu Toivonen, Bernd Gutmann and Kristian Kersting for their many contributions to ProbLog, the Biomine team for the application, and Theofrastos Mantadelis for the development of SimpleCUDD.
This work is partially supported by the GOA project 2008/08 Probabilistic Logic Learning. 
Angelika Kimmig is supported by the Research
Foundation-Flanders (FWO-Vlaanderen).
V\'{\i}tor Santos Costa and
  Ricardo Rocha are partially supported by the research projects
  STAMPA (PTDC/EIA/67738/2006) and JEDI (PTDC/ EIA/66924/2006) and by
  Funda\c{c}\~ao para a Ci\^encia e Tecnologia.

\bibliography{tplp}

\begin{thebibliography}{}

\bibitem[\protect\citeauthoryear{Bachmair, Chen, and Ramakrishnan}{Bachmair
  et~al\mbox{.}}{1993}]{Bachmair-93}
{\sc Bachmair, L.}, {\sc Chen, T.}, {\sc and} {\sc Ramakrishnan, I.~V.} 1993.
\newblock {Associative Commutative Discrimination Nets}.
\newblock In {\em International Joint Conference on Theory and Practice of
  Software Development}. LNCS, vol. 668. Springer, 61--74.

\bibitem[\protect\citeauthoryear{Bryant}{Bryant}{1986}]{Bryant86}
{\sc Bryant, R.~E.} 1986.
\newblock Graph-based algorithms for boolean function manipulation.
\newblock {\em IEEE Trans. Computers\/}~{\em 35,\/}~8, 677--691.

\bibitem[\protect\citeauthoryear{Cussens}{Cussens}{2000}]{Cussens00}
{\sc Cussens, J.} 2000.
\newblock Stochastic logic programs: Sampling, inference and applications.
\newblock In {\em Uncertainty in Artificial Intelligence}, {C.~Boutilier} {and}
  {M.~Goldszmidt}, Eds. Morgan Kaufmann, 115--122.

\bibitem[\protect\citeauthoryear{Dalvi and Suciu}{Dalvi and
  Suciu}{2004}]{DalviS04}
{\sc Dalvi, N.~N.} {\sc and} {\sc Suciu, D.} 2004.
\newblock Efficient query evaluation on probabilistic databases.
\newblock In {\em International Conference on Very Large Databases}, {M.~A.
  Nascimento}, {M.~T. {\"O}zsu}, {D.~Kossmann}, {R.~J. Miller}, {J.~A.
  Blakeley}, {and} {K.~B. Schiefer}, Eds. Morgan Kaufmann, 864--875.

\bibitem[\protect\citeauthoryear{Dantsin}{Dantsin}{1991}]{Dantsin}
{\sc Dantsin, E.} 1991.
\newblock Probabilistic logic programs and their semantics.
\newblock In {\em Russian Conference on Logic Programming}, {A.~Voronkov}, Ed.
  LNCS, vol. 592. Springer, 152--164.

\bibitem[\protect\citeauthoryear{{De Raedt}, Demoen, Fierens, Gutmann,
  Janssens, Kimmig, Landwehr, Mantadelis, Meert, Rocha, {Santos Costa}, Thon,
  and Vennekens}{{De Raedt} et~al\mbox{.}}{2008}]{DeRaedt-NIPSWS08}
{\sc {De Raedt}, L.}, {\sc Demoen, B.}, {\sc Fierens, D.}, {\sc Gutmann, B.},
  {\sc Janssens, G.}, {\sc Kimmig, A.}, {\sc Landwehr, N.}, {\sc Mantadelis,
  T.}, {\sc Meert, W.}, {\sc Rocha, R.}, {\sc {Santos Costa}, V.}, {\sc Thon,
  I.}, {\sc and} {\sc Vennekens, J.} 2008.
\newblock {Towards Digesting the Alphabet-Soup of Statistical Relational
  Learning}.
\newblock In {\em NIPS Workshop on Probabilistic Programming}.

\bibitem[\protect\citeauthoryear{{De Raedt}, Frasconi, Kersting, and
  Muggleton}{{De Raedt} et~al\mbox{.}}{2008}]{DeRaedt-PILPbook}
{\sc {De Raedt}, L.}, {\sc Frasconi, P.}, {\sc Kersting, K.}, {\sc and} {\sc
  Muggleton, S.}, Eds. 2008.
\newblock {\em Probabilistic Inductive Logic Programming - Theory and
  Applications}. LNCS, vol. 4911.

\bibitem[\protect\citeauthoryear{{De Raedt}, Kersting, Kimmig, Revoredo, and
  Toivonen}{{De Raedt} et~al\mbox{.}}{2008}]{DeRaedt08MLJ}
{\sc {De Raedt}, L.}, {\sc Kersting, K.}, {\sc Kimmig, A.}, {\sc Revoredo, K.},
  {\sc and} {\sc Toivonen, H.} 2008.
\newblock Compressing probabilistic {P}rolog programs.
\newblock {\em Machine Learning\/}~{\em 70,\/}~2-3, 151--168.

\bibitem[\protect\citeauthoryear{{De Raedt}, Kimmig, Gutmann, Kersting, {Santos
  Costa}, and Toivonen}{{De Raedt} et~al\mbox{.}}{2009}]{DeRaedt-IQTechReport}
{\sc {De Raedt}, L.}, {\sc Kimmig, A.}, {\sc Gutmann, B.}, {\sc Kersting, K.},
  {\sc {Santos Costa}, V.}, {\sc and} {\sc Toivonen, H.} 2009.
\newblock Probabilistic inductive querying using {ProbLog}.
\newblock Tech. Rep. CW 552, Department of Computer Science, Katholieke
  Universiteit Leuven.

\bibitem[\protect\citeauthoryear{{De Raedt}, Kimmig, and Toivonen}{{De Raedt}
  et~al\mbox{.}}{2007}]{DeRaedt07}
{\sc {De Raedt}, L.}, {\sc Kimmig, A.}, {\sc and} {\sc Toivonen, H.} 2007.
\newblock Prob{L}og: A probabilistic {P}rolog and its application in link
  discovery.
\newblock In {\em International Joint Conference on Artificial Intelligence},
  {M.~M. Veloso}, Ed. 2462--2467.

\bibitem[\protect\citeauthoryear{Fredkin}{Fredkin}{1962}]{Fredkin-62}
{\sc Fredkin, E.} 1962.
\newblock {Trie Memory}.
\newblock {\em Communications of the ACM\/}~{\em 3}, 490--499.

\bibitem[\protect\citeauthoryear{Fuhr}{Fuhr}{2000}]{Fuhr00}
{\sc Fuhr, N.} 2000.
\newblock Probabilistic {D}atalog: Implementing logical information retrieval
  for advanced applications.
\newblock {\em Journal of the American Society for Information Science
  (JASIS)\/}~{\em 51,\/}~2, 95--110.

\bibitem[\protect\citeauthoryear{Getoor and Taskar}{Getoor and
  Taskar}{2007}]{Getoor07}
{\sc Getoor, L.} {\sc and} {\sc Taskar, B.}, Eds. 2007.
\newblock {\em Statistical Relational Learning}.
\newblock The MIT press.

\bibitem[\protect\citeauthoryear{Graf}{Graf}{1996}]{Graf-96}
{\sc Graf, P.} 1996.
\newblock {\em {Term Indexing}}. LNAI, vol. 1053.
\newblock Springer.

\bibitem[\protect\citeauthoryear{Gutmann, Kimmig, Kersting, and {De
  Raedt}}{Gutmann et~al\mbox{.}}{2008}]{Gutmann08}
{\sc Gutmann, B.}, {\sc Kimmig, A.}, {\sc Kersting, K.}, {\sc and} {\sc {De
  Raedt}, L.} 2008.
\newblock Parameter learning in probabilistic databases: A least squares
  approach.
\newblock In {\em European Conference on Machine Learning}, {W.~Daelemans},
  {B.~Goethals}, {and} {K.~Morik}, Eds. LNCS, vol. 5211. Springer, 473--488.

\bibitem[\protect\citeauthoryear{Ishihata, Kameya, Sato, and ichi
  Minato}{Ishihata et~al\mbox{.}}{2008}]{sato:ilp08}
{\sc Ishihata, M.}, {\sc Kameya, Y.}, {\sc Sato, T.}, {\sc and} {\sc ichi
  Minato, S.} 2008.
\newblock Propositionalizing the {EM} algorithm by {BDDs}.
\newblock In {\em Proceedings of Inductive Logic Programming (ILP 2008), Late
  Breaking Papers}, {F.~{\v Z}elezn{\'y}} {and} {N.~Lavra{\v c}}, Eds. Prague,
  Czech Republic, 44--49.

\bibitem[\protect\citeauthoryear{Kersting and {De Raedt}}{Kersting and {De
  Raedt}}{2008}]{Kersting08}
{\sc Kersting, K.} {\sc and} {\sc {De Raedt}, L.} 2008.
\newblock Basic principles of learning bayesian logic programs.
\newblock In {\em Probabilistic Inductive Logic Programming}, {L.~{De Raedt}},
  {P.~Frasconi}, {K.~Kersting}, {and} {S.~Muggleton}, Eds. LNCS, vol. 4911.
  Springer, 189--221.

\bibitem[\protect\citeauthoryear{Kimmig and {De Raedt}}{Kimmig and {De
  Raedt}}{2009}]{Kimmig09}
{\sc Kimmig, A.} {\sc and} {\sc {De Raedt}, L.} 2009.
\newblock Local query mining in a probabilistic {P}rolog.
\newblock In {\em International Joint Conference on Artificial Intelligence},
  {C.~Boutilier}, Ed. 1095--1100.

\bibitem[\protect\citeauthoryear{Kimmig, {De Raedt}, and Toivonen}{Kimmig
  et~al\mbox{.}}{2007}]{Kimmig07}
{\sc Kimmig, A.}, {\sc {De Raedt}, L.}, {\sc and} {\sc Toivonen, H.} 2007.
\newblock Probabilistic explanation based learning.
\newblock In {\em European Conference on Machine Learning}, {J.~N. Kok},
  {J.~Koronacki}, {R.~L. de~M{\'a}ntaras}, {S.~Matwin}, {D.~Mladenic}, {and}
  {A.~Skowron}, Eds. LNCS, vol. 4701. Springer, 176--187.

\bibitem[\protect\citeauthoryear{Kimmig, Gutmann, and {Santos Costa}}{Kimmig
  et~al\mbox{.}}{2009}]{Kimmig-SRL09}
{\sc Kimmig, A.}, {\sc Gutmann, B.}, {\sc and} {\sc {Santos Costa}, V.} 2009.
\newblock Trading memory for answers: Towards tabling {ProbLog}.
\newblock In {\em International Workshop on Statistical Relational Learning}.

\bibitem[\protect\citeauthoryear{Kimmig, {Santos Costa}, Rocha, Demoen, and {De
  Raedt}}{Kimmig et~al\mbox{.}}{2008}]{Kimmig08}
{\sc Kimmig, A.}, {\sc {Santos Costa}, V.}, {\sc Rocha, R.}, {\sc Demoen, B.},
  {\sc and} {\sc {De Raedt}, L.} 2008.
\newblock {On the Efficient Execution of Prob{L}og Programs}.
\newblock In {\em International Conference on Logic Programming}, {M.~G. de~la
  Banda} {and} {E.~Pontelli}, Eds. Number 5366 in LNCS. Springer, 175--189.

\bibitem[\protect\citeauthoryear{Lakshmanan, Leone, Ross, and
  Subrahmanian}{Lakshmanan et~al\mbox{.}}{1997}]{Lakshmanan}
{\sc Lakshmanan, L. V.~S.}, {\sc Leone, N.}, {\sc Ross, R.~B.}, {\sc and} {\sc
  Subrahmanian, V.~S.} 1997.
\newblock Prob{V}iew: A flexible probabilistic database system.
\newblock {\em ACM Transactions on Database Systems\/}~{\em 22,\/}~3, 419--469.

\bibitem[\protect\citeauthoryear{Mantadelis and Janssens}{Mantadelis and
  Janssens}{2009}]{Mantadelis09}
{\sc Mantadelis, T.} {\sc and} {\sc Janssens, G.} 2009.
\newblock Tabling relevant parts of {SLD} proofs for ground goals in a
  probabilistic setting.
\newblock In {\em International Colloquium on Implementation of Constraint and
  LOgic Programming Systems}.

\bibitem[\protect\citeauthoryear{Muggleton}{Muggleton}{1995}]{Muggleton96}
{\sc Muggleton, S.} 1995.
\newblock Stochastic logic programs.
\newblock In {\em Advances in ILP}, {L.~{De Raedt}}, Ed.

\bibitem[\protect\citeauthoryear{Poole}{Poole}{1993a}]{Poole93:jrnl}
{\sc Poole, D.} 1993a.
\newblock Logic {p}rogramming, {a}bduction and {p}robability.
\newblock {\em New {G}eneration {C}omputing\/}~{\em 11}, 377--400.

\bibitem[\protect\citeauthoryear{Poole}{Poole}{1993b}]{Poole:93}
{\sc Poole, D.} 1993b.
\newblock Probabilistic {H}orn abduction and {B}ayesian networks.
\newblock {\em Artificial Intelligence\/}~{\em 64}, 81--129.

\bibitem[\protect\citeauthoryear{Poole}{Poole}{2000}]{Poole00}
{\sc Poole, D.} 2000.
\newblock Abducing through negation as failure: stable models within the
  independent choice logic.
\newblock {\em Journal of Logic Programming\/}~{\em 44,\/}~1-3, 5--35.

\bibitem[\protect\citeauthoryear{Ramakrishnan, Rao, Sagonas, Swift, and
  Warren}{Ramakrishnan et~al\mbox{.}}{1999}]{RamakrishnanIV-99}
{\sc Ramakrishnan, I.~V.}, {\sc Rao, P.}, {\sc Sagonas, K.}, {\sc Swift, T.},
  {\sc and} {\sc Warren, D.~S.} 1999.
\newblock {Efficient Access Mechanisms for Tabled Logic Programs}.
\newblock {\em Journal of Logic Programming\/}~{\em 38,\/}~1, 31--54.

\bibitem[\protect\citeauthoryear{Riguzzi}{Riguzzi}{2007}]{Riguzzi}
{\sc Riguzzi, F.} 2007.
\newblock A top down interpreter for {LPAD} and {CP}-logic.
\newblock In {\em Congress of the Italian Association for Artificial
  Intelligence (AI*IA)}, {R.~Basili} {and} {M.~T. Pazienza}, Eds. LNCS, vol.
  4733. Springer, 109--120.

\bibitem[\protect\citeauthoryear{Santos~Costa}{Santos~Costa}{2007}]{DBLP:conf/%
padl/Costa07}
{\sc Santos~Costa, V.} 2007.
\newblock Prolog performance on larger datasets.
\newblock In {\em Practical Aspects of Declarative Languages, 9th International
  Symposium, PADL 2007, Nice, France, January 14-15, 2007.}, {M.~Hanus}, Ed.
  LNCS, vol. 4354. Springer, 185--199.

\bibitem[\protect\citeauthoryear{{Santos Costa}, Page, Qazi, and
  Cussens}{{Santos Costa} et~al\mbox{.}}{2003}]{Costa03:uai}
{\sc {Santos Costa}, V.}, {\sc Page, D.}, {\sc Qazi, M.}, {\sc and} {\sc
  Cussens, J.} 2003.
\newblock {CLP(BN):} constraint logic programming for probabilistic knowledge.
\newblock In {\em Conference on Uncertainty in Artificial Intelligence},
  {C.~Meek} {and} {U.~Kj{\ae}rulff}, Eds. Morgan Kaufmann, 517--524.

\bibitem[\protect\citeauthoryear{Santos~Costa, Sagonas, and Lopes}{Santos~Costa
  et~al\mbox{.}}{2007}]{jit-index}
{\sc Santos~Costa, V.}, {\sc Sagonas, K.}, {\sc and} {\sc Lopes, R.} 2007.
\newblock Demand-driven indexing of prolog clauses.
\newblock In {\em International Conference on Logic Programming}, {V.~Dahl}
  {and} {I.~Niemel\"a}, Eds. LNCS, vol. 4670. Springer, 305--409.

\bibitem[\protect\citeauthoryear{Sato}{Sato}{1995}]{Sato:95}
{\sc Sato, T.} 1995.
\newblock {A} statistical learning method for logic programs with distribution
  semantics.
\newblock In {\em International Conference on Logic Programming},
  {L.~Sterling}, Ed. MIT Press, 715--729.

\bibitem[\protect\citeauthoryear{Sato and Kameya}{Sato and
  Kameya}{2001}]{SatoKameya:01}
{\sc Sato, T.} {\sc and} {\sc Kameya, Y.} 2001.
\newblock Parameter learning of logic programs for symbolic-statistical
  modeling.
\newblock {\em Journal of Artificial Intelligence Research (JAIR)\/}~{\em 15},
  391--454.

\bibitem[\protect\citeauthoryear{Sevon, Eronen, Hintsanen, Kulovesi, and
  Toivonen}{Sevon et~al\mbox{.}}{2006}]{Sevon06}
{\sc Sevon, P.}, {\sc Eronen, L.}, {\sc Hintsanen, P.}, {\sc Kulovesi, K.},
  {\sc and} {\sc Toivonen, H.} 2006.
\newblock Link discovery in graphs derived from biological databases.
\newblock In {\em Data Integration in the Life Sciences}, {U.~Leser},
  {F.~Naumann}, {and} {B.~A. Eckman}, Eds. LNCS, vol. 4075. Springer, 35--49.

\bibitem[\protect\citeauthoryear{Valiant}{Valiant}{1979}]{Valiant1979}
{\sc Valiant, L.~G.} 1979.
\newblock The complexity of enumeration and reliability problems.
\newblock {\em SIAM Journal on Computing\/}~{\em 8,\/}~3, 410--421.

\bibitem[\protect\citeauthoryear{Vennekens, Verbaeten, and
  Bruynooghe}{Vennekens et~al\mbox{.}}{2004}]{Vennekens}
{\sc Vennekens, J.}, {\sc Verbaeten, S.}, {\sc and} {\sc Bruynooghe, M.} 2004.
\newblock Logic programs with annotated disjunctions.
\newblock In {\em International Conference on Logic Programming}, {B.~Demoen}
  {and} {V.~Lifschitz}, Eds. LNCS, vol. 3132. Springer, 431--445.

\bibitem[\protect\citeauthoryear{Widom}{Widom}{2005}]{Trio}
{\sc Widom, J.} 2005.
\newblock Trio: A system for integrated management of data, accuracy, and
  lineage.
\newblock In {\em Conference on Innovative Data Systems Research}. 262--276.

\end{thebibliography}

\end{document}